\documentclass[apj]{emulateapj}
\usepackage[colorlinks,linkcolor={blue},citecolor={blue},urlcolor={red}]{hyperref}
\usepackage{bm}
\usepackage{graphicx}
\usepackage{epsf}
\usepackage{graphics}
\usepackage{amsmath}
\usepackage{lineno}

\def\beq{\begin{equation}}
\def\eeq{\end{equation}}
\def\bey{\begin{eqnarray}}
\def\eey{\end{eqnarray}}

\def\Mpc{\,{\rm Mpc}}
\def\mpc{\, h^{-1}{\rm {Mpc}}}

\def\kpc{\, h^{-1}{\rm {kpc}}}

\def\kms{\,{\rm {km\, s^{-1}}}}

\def\msun{\, h^{-1}{\rm M_\odot}}

\def\gs{\mathrel{\raise1.16pt\hbox{$>$}\kern-7.0pt
\lower3.06pt\hbox{{$\scriptstyle \sim$}}}}
\def\ls{\mathrel{\raise1.16pt\hbox{$<$}\kern-7.0pt
\lower3.06pt\hbox{{$\scriptstyle \sim$}}}}
\def\gtsima{\, {\buildrel > \over \sim} \,}
\def\ltsima{\, {\buildrel < \over \sim} \,}
\def\prosima{\, {\buildrel \propto \over \sim} \,}
\def\gsim{\lower.5ex\hbox{\gtsima}}
\def\lsim{\lower.5ex\hbox{\ltsima}}
\def\simgt{\lower.5ex\hbox{\gtsima}}
\def\simlt{\lower.5ex\hbox{\ltsima}}
\def\simpr{\lower.5ex\hbox{\prosima}}

\shorttitle{Gas in Cosmic Web}
\shortauthors{R.J. Li et al.}

\begin{document}
\title{ELUCID VII: Using constrained hydro 
simulations to explore the gas component of 
the cosmic web}

\author{Renjie Li\altaffilmark{1,2}, Huiyuan Wang\altaffilmark{1,2}, H. J. Mo\altaffilmark{3}, Shuiyao Huang\altaffilmark{3}, Neal Katz\altaffilmark{3}, Xiong Luo\altaffilmark{1,2}, Weiguang Cui\altaffilmark{4,5}, Hao Li\altaffilmark{1,2}, Xiaohu Yang\altaffilmark{6,7}, Ning Jiang\altaffilmark{1,2}, and  Yuning Zhang\altaffilmark{8}}
\altaffiltext{1}{Key Laboratory for Research in Galaxies and Cosmology, Department of Astronomy, University of Science and Technology of China, Hefei, Anhui 230026, China;phylrj@mail.ustc.edu.cn, whywang@ustc.edu.cn}
\altaffiltext{2}{School of Astronomy and Space Science, University of Science and Technology of China, Hefei 230026, China}
\altaffiltext{3}{Department of Astronomy, University of Massachusetts, Amherst MA 01003-9305, USA}
\altaffiltext{4}{Institute for Astronomy, University of Edinburgh, Royal Observatory, Edinburgh EH9 3HJ, United Kingdom}
\altaffiltext{5}{Departamento de Física Teórica, M-8, Universidad Autónoma de Madrid, Cantoblanco 28049, Madrid, Spain}
\altaffiltext{6}{Department of Astronomy, School of Physics and Astronomy, and Shanghai Key Laboratory for Particle Physics and Cosmology, Shanghai Jiao Tong University, Shanghai 200240, China}
\altaffiltext{7}{Tsung-Dao Lee Institute and Key Laboratory for Particle Physics, Astrophysics and Cosmology, Ministry of Education,  Shanghai Jiao Tong University, Shanghai 200240, China}
\altaffiltext{8}{Department of Astronomy, Tsinghua University, Beijing 100084, China}
\begin{abstract}
Using reconstructed initial conditions in the SDSS survey volume,    
we carry out constrained hydrodynamic simulations in three regions 
representing different types of the cosmic web: the Coma cluster of galaxies;
the SDSS great wall; and a large low-density region at $z\sim 0.05$. 
These simulations, which include star formation and stellar feedback but 
no AGN formation and feedback, are used to investigate the properties and 
evolution of intergalactic and intra-cluster media. About half of the warm-hot 
intergalactic gas is associated with filaments in the local cosmic 
web. Gas in the outskirts of massive filaments and halos can be heated 
significantly by accretion shocks generated by mergers of filaments and 
halos, respectively, and there is a tight correlation between gas 
temperature and the strength of the local tidal field. The simulations also 
predict some discontinuities associated with shock fronts and contact 
edges, which can be tested using observations of the thermal SZ effect and X-rays.
A large fraction of the sky is covered by Ly$\alpha$ and OVI absorption
systems, and most of the OVI systems and low-column density HI systems are associated with filaments in the cosmic web. 
The constrained simulations, which follow the formation and 
heating history of the observed cosmic web, provide an important avenue to 
interpret observational data. With full information about the origin 
and location of the cosmic gas to be observed, such simulations can 
also be used to develop observational strategies. 
\end{abstract}
\keywords{galaxies: halos - galaxies: general -- methods: observational - methods: statistical}

\section{Introduction}
\label{sec_intro}

In the local Universe, only about ten percent of the baryons are found to be locked in stars 
and dense interstellar media
\citep[e.g.][]{Bregman2007}. Most of the baryons are, therefore, expected to be ionized and diffuse in various components of the cosmic web, from virialized dark matter halos to large-scale filaments and sheets. The thermal, dynamical, and chemical states of the baryonic gas are subject to   
a number of processes, such as gravitational collapse, radiative heating/cooling and 
feedback from stars and AGNs, all closely related to the formation and evolution 
of the objects we observe. Clearly, the study of the gas components in the cosmic web 
is a key step to understanding the formation and evolution of galaxies and the
large-scale structure of the universe.  

Observationally, the gas components can be investigated using various methods,
such as X-ray emission, Sunyaev-Zel’dovich (SZ) effects, and emission
and absorption lines. The intra-cluster medium (ICM), which is 
the hottest component in the cosmic web, can be observed 
as extended X-ray sources \citep[e.g.][]{Mirakhor2020MNRAS, Churazov2021A&A}.
The free electrons in the hot ICM can also change the energy distribution of the 
CMB photons via inverse Compton scattering, producing a thermal SZ 
effect \citep[tSZ,][]{SZ1972CoASP} that can be used to investigate 
the thermal content of the ICM \citep[e.g.][]{Planck2013A&AV, Bleem2015ApJS}. 
Since the X-ray emission and tSZ effect depend on gas density and temperature 
in different ways, the combination of X-ray and tSZ observations can provide 
measurements that can be used to infer the thermodynamic properties   
of the ICM \citep[e.g.][]{Ghirardini2018A&A, Mirakhor2020MNRAS, Churazov2021A&A}.
The gas in smaller and more abundant groups is colder 
and thus more difficult to detect individually. A common practice 
is to adopt some stacking technique and investigate the observational signal statistically 
\citep[e.g.][]{Planck2013A&AXI, WangLei2014MNRAS, Ma2015JCAP, Lim2018ApJ}. 

The baryonic gas in filaments and sheets is more diffuse and colder, and is
thus even harder to investigate  using X-ray emission and the tSZ effect. 
Significant detections of the tSZ signal and X-ray emission has been 
obtained only for massive filaments between pairs of massive 
merging clusters \citep[e.g.][]{Planck2013VIII, Sugawara2017PASJ, Reiprich2021A&A}. 
Faint X-ray filaments around individual clusters 
are detected for a few clusters with extremely deep observations
\citep{Eckert2015Natur, Connor2018ApJ}. By stacking a large 
number of galaxy pairs, one may measure the weak tSZ effect 
produced by the gas between galaxies \citep[e.g.][]{deGraaff2019}. 
However, the interpretation of the stacking results is not 
straightforward, because it is unclear whether the signal is 
dominated by the diffuse gas associated with the filaments
or by small halos embedded in the filaments. 

The gas components can also be observed through absorption 
lines that they produce in the spectra of background sources like quasars, 
such as Ly$\alpha$, MgII, and OVI absorption line systems 
\citep[e.g.][]{Penton2000ApJS, Tumlinson2011Sci, Shull2012ApJ, Rudie2012ApJ, 
Werk2013ApJS,  Danforth2016}. 
Some of the observed absorption lines are clearly associated with galaxies, 
indicating that they are caused by the gas around galaxies (gaseous halos) 
or even by the gas associated with outflows driven by stellar and AGN feedback
\citep[e.g.][]{Rudie2013ApJ, Lan2018ApJ}. Filament gas that is not physically 
connected to any halos can also produce observable absorption line 
systems \citep[e.g.][]{Pessa2018MNRAS, Nicastro2018Natur}. 
Other methods, such as emission lines and radio emission, have also 
been used to study gas in galaxy groups and clusters 
\citep[see e.g.][]{Brown2011MNRAS, Zhang2016ApJ, Gheller2020MNRAS}.
Despite all these, a large fraction of the gas, about 30 to 40 percent of the 
baryons, remain undetected in the low-redshift universe 
\citep[e.g.][]{Bregman2007,Shull2012ApJ}, and the details of the 
physical processes responsible for the evolution and 
state of the gas is still unclear.

Theoretically, numerical simulations play a critical role in 
understanding the physical processes that determine gas properties 
in the cosmic web. Hydrodynamic simulations show that the gas in the cosmic web 
is processed and heated by strong shocks during the formation of halos and filaments \citep[][]{Keres2005MNRAS, Dekel2006MNRAS, Kang2007ApJ, Bykov2008SSRv, Schaal2015MNRAS, Zinger2018MNRAS}. 
In particular, more than 40 percent of the baryonic gas at $z\sim 0$  
is found to be located outside halos, in a Warm-Hot Intergalactic Medium (WHIM)
with a temperature between $10^5$ and $10^7{\rm K}$ 
\citep[e.g.][]{Cen1999ApJ, Dave2001ApJ, Haider2016MNRAS, Cui2019MNRAS, Martizzi2019MNRAS}. 
Most of the gas in the WHIM tends to reside in filaments, 
and the relatively low density and high temperature makes the gas   
hard to detect in both emission and absorption. 
Simulations also show that AGN feedback, stellar winds and ram pressure 
stripping can not only affect the thermodynamic properties 
and spatial distribution of the gas in the 
cosmic web \citep[see e.g.][]{SunM2009, Lovisari2015, Lim2018MNRAS, Truong2021MNRAS, Amodeo2021, Schaan2021}, 
but also chemically enrich the circumgalactic medium (CGM) and intergalactic 
medium (IGM) \citep[e.g.][]{Cen2006ApJ, Liang2016MNRAS, Oppenheimer2016MNRAS, 
Rahmati2016MNRAS, Nelson2018MNRAS, Boselli2021arXiv}. 

\begin{figure*}
    \centering
    \includegraphics[scale=0.85]{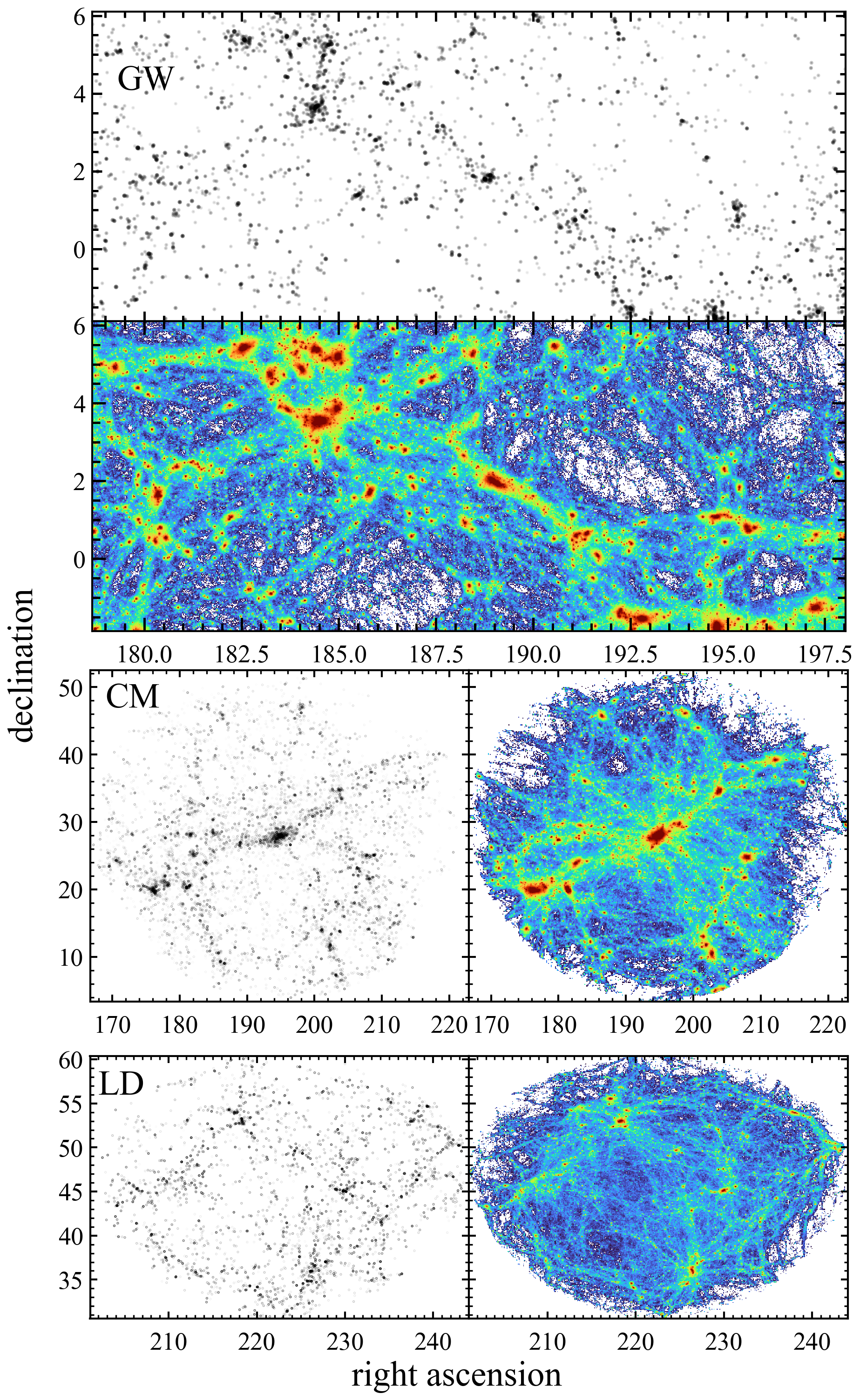}
	\caption{These six panels show the spatial distribution of galaxies (black dots) and dark matter (color-coded maps) in the constrained simulations of the Great Wall region (GW, top two panels), the Coma cluster region (CM, middle two panels), and the low density void region (LD, bottom two panels) in the J2000.0 right ascension and declination. Note that the upper panel of GW and the left panels of CM and LD show real galaxies from the SDSS galaxy redshift survey. 
	}\label{fig_galaxydm}
\end{figure*}

Clearly, to constrain models using observations,  
it is necessary to compare model predictions and observational data in a 
meaningful way. In general, the comparison is made in a statistical sense, 
in terms of summary statistics, such as distribution functions and 
correlations. Since both observational and simulation samples 
are finite, one needs to deal with the cosmic variance 
to achieve an unbiased comparison between models and data.
This is a serious challenge for both observations and simulations. 
For example, large-scale structures in the cosmic web, such as filaments and 
massive clusters are rare, so that the observational samples 
are usually small; observations of absorption systems are limited 
by the background sources available, so that only the gas distribution
along a limited number of lines of sight can be sampled. All these 
make it difficult to obtain a statistically fair sample for the 
objects concerned. Hydrodynamic simulations are limited by the 
dynamic range they cover, and the simulation volume is usually relatively 
small, making it difficult to sample fairly the large-scale structure of 
the cosmic web. It is thus imperative to have as much theoretical 
and empirical input as possible to help comparing model predictions 
with the limited amount of observational data in an unbiased way. 
The bias can be minimized if comparisons are made for systems that have 
both the same environments and the same formation histories. 
This can be achieved if one can accurately reconstruct the initial conditions 
for the formation of the structures from which the actual gas emission and 
absorption are produced. Such a reconstruction is now possible
\citep[e.g.][]{Hoffman1991ApJ, Nusser1992ApJ, Frisch2002Natur, Klypin2003ApJ, Jasche2013MNRAS, Kitaura2013MNRAS, WangH2014ApJ, Sorce2016MNRAS, Horowitz2019ApJ, Modi2019JCAP, Bos2019MNRAS,Kitaura2021MNRAS}. As shown in \citet[][]{WangH2016ApJ}, 
constrained simulations using such reconstructed initial conditions can 
provide reliable representations of the structure, motion,  
and formation history of the objects observed in the real universe. 
Hydrodynamic simulations can also be carried out with these reconstructed 
initial conditions to follow the evolution of the gas component, 
facilitating unbiased comparisons between model predictions and 
observational data. Constrained simulations can also provide reliable
tracers of the mass distribution in the observed universe,
which is important for stacking analyses to discover unidentified gas in the cosmic 
web \citep[e.g.][]{Lim2018MNRAS, Lim2020ApJ}.

In this paper, we use hydrodynamic simulations to follow the evolution of the  
gas component in three volumes covered by the Sloan Digital Sky Survey
\citep[SDSS;][]{York2000AJ}, using initial conditions reconstructed from the 
ELUCID project \citep{WangH2014ApJ, WangH2016ApJ}. The three volumes 
are chosen to contain the Coma cluster, SDSS great wall, and a low-density 
`void'. The main goal of this paper is to demonstrate the 
power and uniqueness of constrained simulations in investigating 
and understanding the gas component in the observed cosmic web.
The paper is organized as follows. Section \ref{sec_cs} describes the three 
constrained hydrodynamic simulations and the method to characterize the cosmic web. 
Section \ref{sec_evo} shows the evolution of gas properties in and 
around these large-scale structures in the local Universe. 
We analyze the properties of the gas and study how they are correlated 
with the local properties of the cosmic web in Section \ref{sec_gascor}. 
We discuss how to use constrained simulations to facilitate 
unbiased comparisons between model predictions and observational data  
in Section \ref{sec_obs}. Finally, we summarize and discuss our results in Section \ref{sec_sum}.

\section{Constrained simulations of the local universe}\label{sec_cs}

\subsection{Initial conditions}

To simulate the evolution of the large scale structure observed in the local Universe, 
we use initial conditions reconstructed from the ELUCID project in the volume 
covered by the SDSS galaxy redshift survey. The reconstruction consisted of four steps.
First, we identified galaxy groups from the SDSS galaxy catalog \citep{Yang2005MNRAS,Yang2007ApJ}. 
Second, we corrected the distance of each group for redshift distortions 
using linear perturbation theory \citep{WangH2012MNRAS}. Third, we reconstructed 
the present-day mass density field using a halo-domain method 
\citep{WangH2009MNRAS,WangH2016ApJ}. Finally, we employed an HMC$+$PM technique to 
recover the initial conditions from the present-day density field \citep{WangH2013ApJ,WangH2014ApJ}. 
\cite{WangH2016ApJ} performed a large $N$-body simulation using these reconstructed 
initial conditions in the SDSS Dr7 \citep{Abazajian2009ApJS} survey volume 
at $z\le 0.12$. In what follows, we refer to this constrained simulation (CS) as the 
original CS (hereafter OCS). A comprehensive description of these steps and the OCS can be found in \cite{WangH2016ApJ}, and we refer readers to that paper for details. 

In this paper, we focus on three regions 
that contain special large-scale structures of interest. We adopt a 
zoom-in technique, in which the interested structures are simulated with 
a higher resolution than other regions in the simulation volume \citep{Katz1993ApJ}. 
To do this, we select, for each zoom-in simulation, a high 
resolution region (HIR) to contain the particular structure of interest
from the $z=0$ snapshot of the OCS. The HIR is either a cuboid or a spherical 
volume (see below and Table \ref{tab_HIR}). We follow all the particles 
in the HIR back to an initial time. 
We then select an initial HIR (hereafter iHIR) that has exactly the same comoving center, 
shape and orientation (for a cuboid geometry) as 
the HIR. The iHIR is set to contain all particles in the HIR, and so 
the comoving size of the iHIR is larger than that of the HIR. To 
prevent lower-resolution particles from entering the HIR region 
in the subsequent evolution, a buffer region is used in defining the iHIR.
The size of the buffer region is set to be 5\% of the iHIR, so that the iHIR volume 
is enlarged by 15.8\%.

The iHIR is sampled with high-resolution particles, and the 
cosmic density field outside the iHIR is sampled with particles 
of three successively lower resolutions. Except for the lowest resolution 
region, the other two low-resolution regions have exactly the same geometry 
and center as the iHIR. The size of the first low-resolution region 
(hereafter LIR1) is twice that of the iHIR, and its mass resolution is 8 times lower
so that the number of LIR1 particles is similar to that in the iHIR. 
The size and mass resolution of the second low-resolution region (LIR2) are chosen 
to be twice as large and 8 times as low as those of the LIR1, respectively. 
Finally, the rest of the simulation volume is sampled with
particles, each with a mass that is eight times as large as that of 
a LIR2 particle. 
To set up the zoom-in initial condition, we generate a displacement field for high resolution particles with a grid of $6144^3$ cells in the whole simulation box. 
In the three lower resolution regions, we bin the high resolution particles
according to the corresponding mass resolution. 
Only high-resolution regions contain gas particles. We thus split
each high-resolution particle into one dark matter particle and one gas particle. The mass ratio of the two particles equals to $(\Omega_{\rm m,0}-\Omega_{\rm b,0})/\Omega_{\rm b,0}$ and their separation is half of the mean separation of particles. Our inspection shows that this choice of resolution hierarchy 
ensures that no lower-resolution particle enters the HIR region.  
For reference, we list the particle numbers used in iHIR, LIR1 and LIR2 
in Table \ref{tab_HIR}. 

\begin{table*}
    \centering
    \caption{The properties of the three simulated HIR. The table contains the location of the HIR center ($\alpha_{\rm J}$, $\delta_{\rm J}$, and redshift $z$), geometry, comoving size, and mean density ($\rho_{\rm m}$, in units of the cosmic mean density). $N_{\rm g}$ and $N_{\rm d1}$ are the numbers of gas and dark matter particles in the iHIR, $N_{\rm d2}$ and $N_{\rm d3}$ are the numbers of dark matter particles in the LIR1 and LIR2 regions, respectively. The sky coverage of the three CSs are shown in Figure \ref{fig_galaxydm}.}
    \begin{tabular}{c c c c c c c c c c c}
        \hline\
        Simulation & $\alpha_{\rm J}$ & $\delta_{\rm J}$ & $z$ & Geometry & size & $\rho_{\rm m}$ & $N_{\rm g}$ & $N_{\rm d1}$ & $N_{\rm d2}$ & $N_{\rm d3}$\\
        \hline
        CM & 194.8 & 27.9 & 0.0241 & sphere  & $R=30\mpc$ & 1.39 & 7.04e8 & 7.04e8 & 6.17e8 & 6.18e8 \\
        GW & 188.0 & 2.0 & 0.0775  & cuboid & $75\times30\times30(\mpc)^3$ & 2.19 & 9.91e8 & 9.91e8 & 8.67e8 & 8.66e8\\
        LD & 222.6 & 45.5 & 0.0519  & sphere & $R=40\mpc$ & 0.57 & 1.09e9 & 1.09e9 & 9.56e8 & 9.56e8\\
        \hline
    \end{tabular}
    \label{tab_HIR}
\end{table*}

The three HIRs that we choose to simulate are the following. 
The first one contains a part of the SDSS Great Wall  
at redshift $\sim 0.08$. Since the Great Wall is a long filamentary structure, 
we choose a cuboid HIR with size $75\times30\times30(\mpc)^3$. Hereafter, 
we refer to this zoom simulation as the GW simulation. The second one 
is centered on the Coma galaxy cluster at a redshift of 0.0241. The HIR is set to be 
a spherical volume with a radius of $30\Mpc/h$. We refer to the corresponding 
zoom simulation as the CM simulation. The third one is a low-density 
region at $z\sim0.05$, and is chosen to be a spherical volume
with a radius of $40\Mpc/h$. We refer to this simulation as the LD simulation. 
Figure \ref{fig_galaxydm} shows the location of the three selected regions in 
the J2000.0-coordinate system represented by right ascension 
($\alpha_{\rm J}$) and declination ($\delta_{\rm J}$).  
The mean mass densities in the three HIRs are 2.19 (GW), 
1.39 (CM) and 0.57 (LD) times the cosmic mean density. 
In Table \ref{tab_HIR}, we list the basic information of the three HIRs.

We adopt cosmological parameters from WMAP5 
\citep{Dunkley2009ApJS}, the same as the OCS: $\Omega_{\Lambda,0}$ = 0.742, $\Omega_{m,0}$ = 0.258, $\Omega_{b,0}$ = 0.044, h = $\rm H_0/100\kms\Mpc$ = 0.72, $\sigma_{8}$ = 0.80.
The mass resolution of the high-resolution regions are 8 times higher than 
the OCS simulation. The corresponding masses of dark matter and gas particles in the HIR are
$3.20\times10^7\msun$ and $6.74\times10^6\msun$, respectively. 
The mass resolution of our HIRs is close to that used in \cite{Huang2020MNRAS}. 
The masses of
dark matter particles in LIR1, LIR2 and the rest part of the simulation are $3.09\times10^8$, $2.47\times10^9$ and $1.98\times10^{10}$ $\msun$, respectively.
All the three simulations follow the evolution of the cosmological 
density field in a periodic box of comoving length $500\mpc$, and the 
initial redshift of each simulation is set at $z=120$. 
Softening lengths for high resolution particles are all set 
to be 1.8 comoving $\kpc$.

\subsection{Simulation code}

We run all the simulations using Gadget-3, an updated version of Gadget-2
\citep{Springel2005MNRAS}, as described in \cite{Huang2019MNRAS, Huang2020MNRAS}. The gravitational
forces are evaluated using a particle mesh and oct-tree algorithm.
The code includes several recent numerical improvements in the SPH technique
\citep{Huang2019MNRAS}. To summarise, we use the pressure-entropy formulation
\citep{Hopkins2013MNRAS} of SPH to integrate the fluid equations and a quintic spline
kernel to measure fluid quantities over 128 neighbouring particles. We also use
the \citet{Cullen2010MNRAS} viscosity algorithm and artificial conduction as in
\citet{Read2012MNRAS} to capture shocks more accurately and to reduce numerical noise.
Both the artificial viscosity and the conduction are turned on
only in converging flows with $\nabla \cdot \mathbf{v} < 0$ to minimise
unwanted numerical dissipation. We also include the Hubble flow while
calculating the velocity divergence. Our fiducial code leads to considerable
improvements in resolving the instabilities at fluid interfaces in subsonic
flows and produces consistent results with other state-of-art hydrodynamic
codes in various numerical tests \citep{Sembolini2016a, Sembolini2016b, Huang2019MNRAS}.

Besides including cooling from hydrogen and helium and the effects of
a photoionising UV background field \citep{HM12},
we also add metal line cooling including
photoionisation effects for 11 elements as in \citet{Wiersma2009MNRAS}, and we
recalculate cooling rates according to the ionising background
\citep{HM12}. The star formation processes are modelled as in
\citet{Springel2003MNRAS}, which includes a subgrid model for the multiphase interstellar medium (ISM) in
dense regions with $n_\mathrm{H} > 0.13\ \mathrm{cm^{-3}}$, and a star
formation recipe that is scaled to match the Kennicutt-Schmidt relation. In
this paper we will distinguish SPH particles as galaxy particles
based on whether or not their densities are higher than this
density threshold. We specifically trace the enrichment of four metal species C,
O, Si, Fe that are produced from type II SNe, type Ia SNe and AGB stars as in
\citet{Oppenheimer2008MNRAS}. These processes also generate energy that we add to the
simulations as thermal energy.  However, the input energy from these
feedback processes only have sub-dominant effects to galaxy formation compared
to the wind feedback \citep{Oppenheimer2008MNRAS}, as it is typically quickly radiated
away.

As in \cite{Huang2019MNRAS,Huang2020MNRAS}, we adopt a kinetic sub-grid model for the
stellar feedback.
We implemented the new wind algorithm into our SPH code based on
\textsc{gadget-3} (see \citet{Springel2005MNRAS} for reference).
SPH particles in star-forming regions have some probability
of being ejected from their host galaxy in the direction $\vec v\times\vec a$,
implemented with a one-at-a-time
particle ejection algorithm. We temporarily shut off the hydrodynamic forces
to allow particles to escape the dense ISM.
While this is artificial, it simply recognises
the fact that feedback {\it does} drive outflows in real galaxies
(and in ultra-high resolution simulations of isolated galaxies,
like those of \cite{Hopkins2012MNRAS}),
rather than being locally thermalized and radiated away,
owing to collective effects in a multi-phase ISM.

We adopt a variant
of the wind algorithm implemented in GADGET,  using momentum driven
wind models very similar to those found in high-resolution zoom
simulations.  Building on the work of \cite{Oppenheimer2006MNRAS},
we model the launch of galactic winds from star forming
galaxies with controlled parameters of the mass loading factor, $\eta$
$\equiv$ ejection rate/SFR
($\eta\sim\sigma_{gal}^{-\beta}$ for small $\sigma_{gal}$ and
$\eta\sim\sigma_{gal}^{-\alpha}$ for large $\sigma_{gal}$)
and the initial wind speed, $v_{w}$, that scales linearly with the
velocity dispersion of the galaxy ($\sigma_{gal}$). We identify
galaxies on-the-fly at intermittent time-steps during a simulation using a
friend-of-friend (FoF) group finder, which at the same time computes the
properties of these galaxies and estimates the dispersion using the total mass
of the galaxy.

We adopt the same set of wind
parameters as the fiducial simulation from \cite{Huang2020MNRAS},
which was the most successful wind launch scalings, in terms of matching a
broad range of observations \citep[e.g.][]{Dave2013MNRAS}, where
$\alpha=1$ and $\beta=3$, i.e. $\sim$momentum driven wind scalings
for large $\sigma_{gal}$ and supernova-energy driven wind scalings for
small $\sigma_{gal}$, and $V_{w}\sim\sigma_{gal}$.
These scalings are very similar to those found in very high resolution galaxy
zoom simulations \citep{Hopkins2012MNRAS, Hopkins.keres2014,Muratov2015MNRAS}.  Our wind scalings are at wind launch from the star-forming
regions of the galaxy while the very high resolution zoom simulations report
their wind scalings at $0.25 r_{\rm vir}$ \citep{Muratov2015MNRAS}.
Hence, we have had to slightly increase our wind launch velocities to reproduce
their behaviour at $0.25 r_{\rm vir}$.  This seemingly minor change has
important effects; if we were to simply apply the \cite{Muratov2015MNRAS}
scalings at launch, as in all past work, then owing to gravitational
deceleration many wind particles in moderately large galaxies would not even
reach $0.25 r_{\rm vir}$ and those that do would have a $V_w$ that is
almost independent of $\sigma_{gal}$.
We also cap the wind speed so that the energy in the winds
does not exceed that available in supernova.

Feedback from active galactic nuclei (AGNs)
is not included in our simulations; we will discuss the potential impact
of not including AGNs later. Since we are mainly interested in gas
properties at large scales, the impact of AGN feedback may be
relatively small.

\begin{figure*}
    \centering
    \includegraphics[scale=0.5]{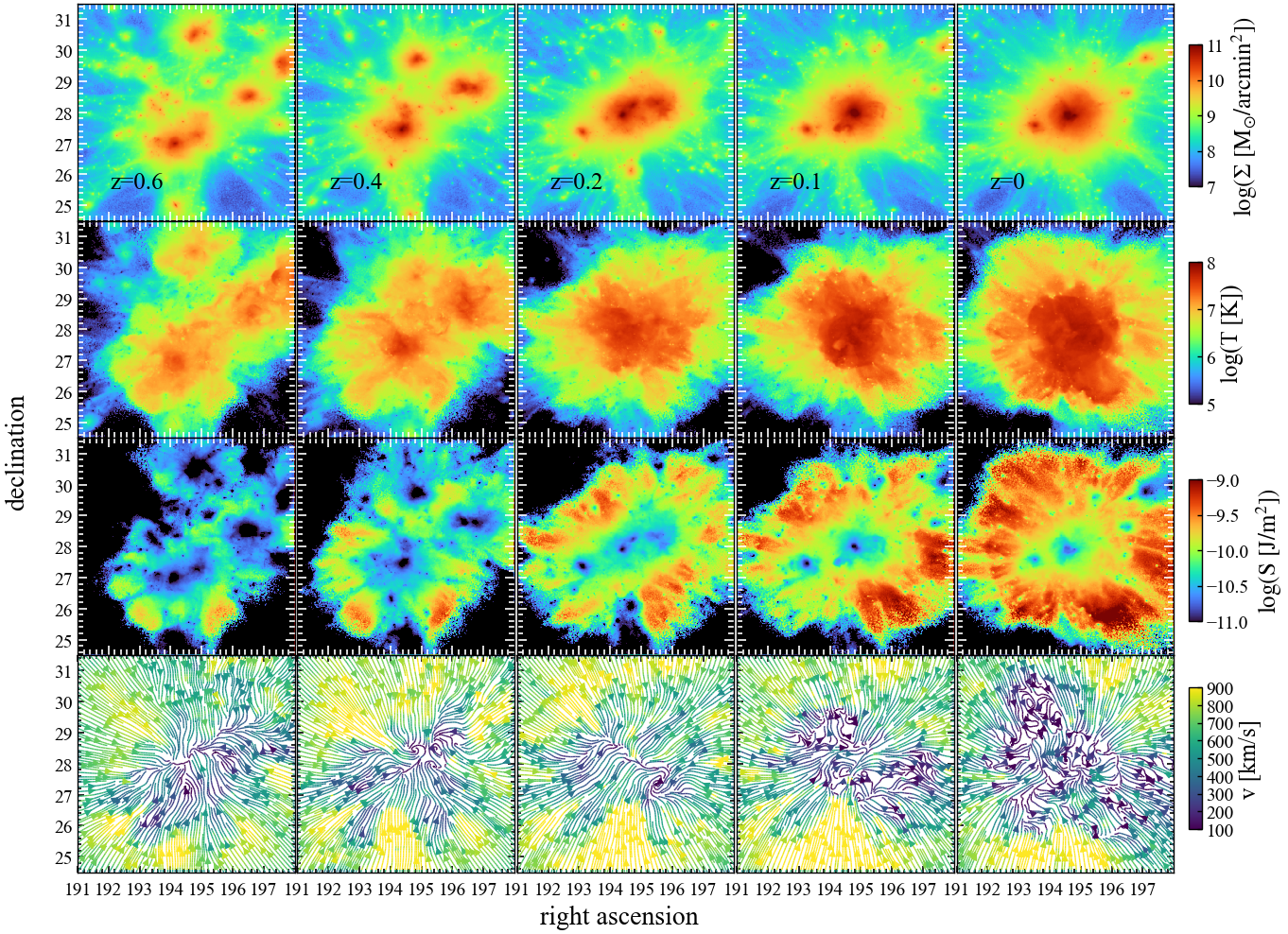}
	\caption{The evolution of the Coma galaxy cluster in the CM simulation from $z=0.6$ to $z=0$. The top shows the surface mass density of gas particles, the second row shows the gas temperature, the third row shows the gas entropy, and the bottom row shows the velocity component that is perpendicular to the line of sight. Note that the mean velocity is removed.}\label{fig_comaev}
\end{figure*}

\subsection{Characterizing the Cosmic Web}\label{sec_tweb}

Halos are identified using a friends-of-friends (FoF) algorithm \citep{Davis85} 
with a linking length $b$ equal to 0.2 times the mean dark matter particle 
separation. The FoF algorithm is first applied to high-resolution DM particles to
identify dark matter (DM) halos. A gas or star particle is assigned to the 
same halo defined by the DM particles if its distance to the nearest 
DM particle is less than $b$. Halos with at least 20 DM particles are identified, 
and the halo mass, $M_{\rm h}$, is defined as the sum of the masses of all 
particles in the halo. Galaxies, including star and star-forming gas particles, 
are identified using the SKID (Spline Kernel Interpolative Denmax) 
algorithm \citep{Keres2005MNRAS}. 

In addition to halos, we are also interested in other components of the cosmic web.
We adopt the so-called ``T-Web'' method \citep[e.g.][]{Hahn2007} to
classify the cosmic web. The method is based on the eigenvalues of
the tidal tensor defined as
\begin{equation}\label{eq_tij}
{\cal T}_{ij}=\partial_i\partial_j\phi\,,
\end{equation}
where, $\phi$ is the peculiar gravitational potential 
and obeys a modified Poisson equation:
\begin{equation}
\nabla^2\phi=\delta
\label{eq_phi}\,.
\end{equation}
Note that this equation is scaled by $4\pi G\bar{\rho}$ and $\delta$ is the 
mass overdensity defined over grid cells and smoothed on a comoving scale 
of $R_{\rm s}$. The three eigenvalues of ${\cal T}_{ij}$ are denoted by 
$\lambda_i$, where $i=1,2$ and 3, with $\lambda_1\ge\lambda_2\ge\lambda_3$. 
Following common practice, we define knots as locations where all three 
eigenvalues are above a given threshold value $\lambda_{\rm th}$, 
while filaments, sheets and voids are locations where   
two, one and none of the engenvalues are above $\lambda_{\rm th}$, respectively.

We follow \cite{Hahn2007} and adopt $\lambda_{\rm th}=0$. Since we want to classify 
the cosmic web on small scales, we choose a smoothing scale of 
$R_{\rm s}=0.5\mpc$, much smaller than the $2.0\mpc$ often adopted 
in previous studies. We will discuss the impact of using different 
smoothing scales. Thus, the knots defined here include not only massive 
galaxy clusters but also small halos; only the most underdense regions 
are classified as voids, as we will see below. 

To distinguish gas in different components of the cosmic web, 
we refer the gas in halos as halo gas, and the gas outside halos as diffuse gas. 
The diffuse gas is further separated into knot gas, filament gas, sheet gas,
and void gas. In the following analysis, we exclude gas particles within galaxies, defined as gas that has $\rm n>0.13~cm^{-3}$ and a temperature less than $\rm 10^{4.5}K$, 
as we are only interested in gas properties on larger scales.

\section{Evolution of large-scale structures in the local Universe}\label{sec_evo}

\begin{figure*}
    \centering
    \includegraphics[scale=0.45]{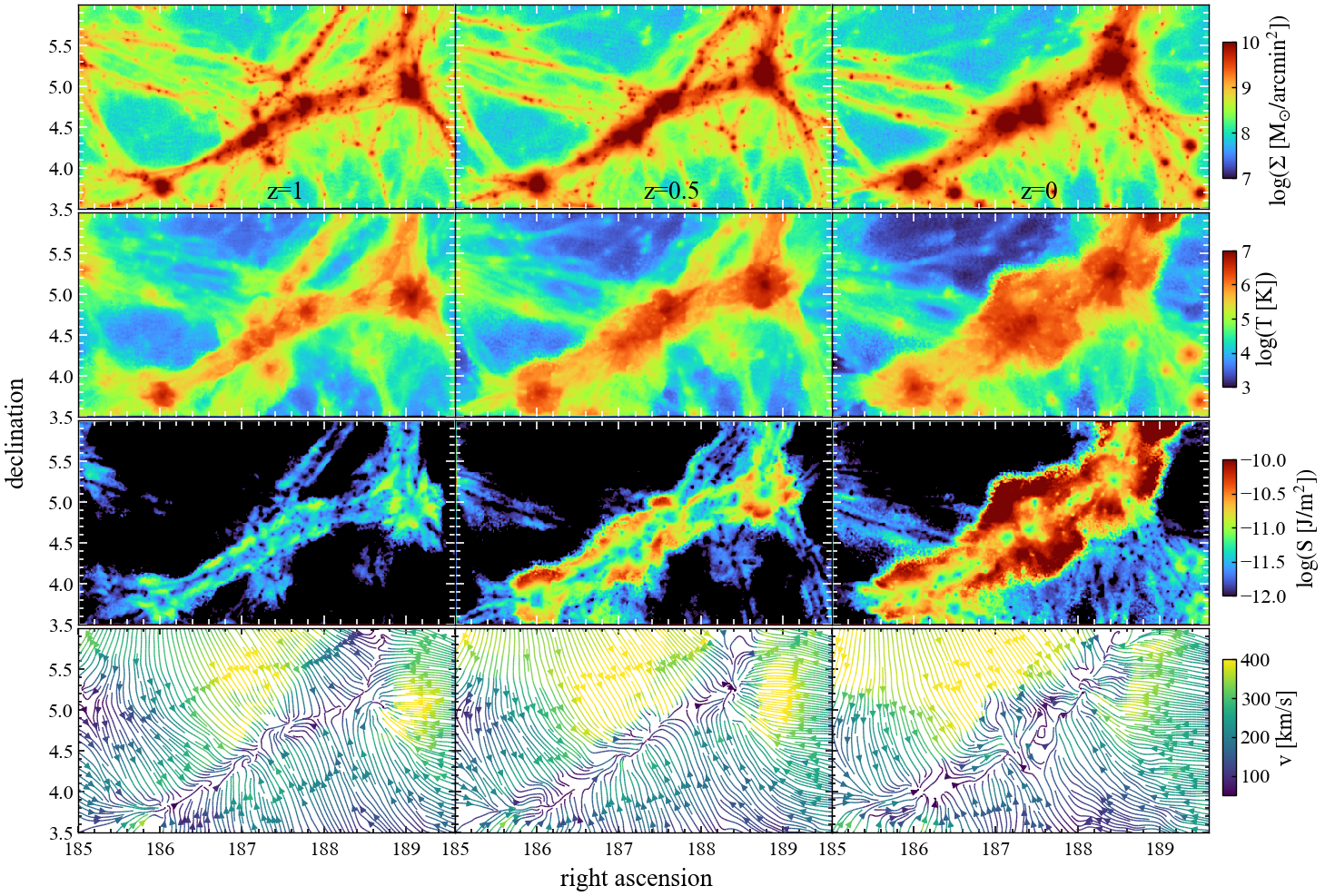}
	\caption{Similar to Figure \ref{fig_comaev} but for a massive filament in the GW simulation. Here we present the results from $z=1$ to 0. Note that this filament is only a small part
of the SDSS Great Wall.
	}\label{fig_gwev}
\end{figure*}

 Figure \ref{fig_galaxydm} shows the spatial distribution of SDSS galaxies 
in the three selected regions. Galaxies are viewed from Earth and shown in 
the J2000.0-coordinate system represented by the right ascension 
($\alpha_{\rm J}$) and declination ($\delta_{\rm J}$).   
The present-day density fields of the corresponding constrained simulations
are also plotted for comparison. As one can see, most of the massive
clusters and filaments and large voids are well reproduced in the constrained 
simulations, as well as some small filaments. We refer readers to \cite{WangH2016ApJ} 
for a detailed discussion about the statistics of the original constrained 
simulation and its comparison with the distributions of real galaxies.
In what follows, we focus on the spatial distribution of gas properties, such as
mass, temperature, entropy ($s\equiv kT/n^{2/3}$) and velocity, as well as
the time evolution of these properties. 

\subsection{The Coma galaxy cluster (CM) simulation}

We first show the evolution of the gas in a region around the 
Coma galaxy cluster (Figure \ref{fig_comaev}). As one can see, the Coma cluster has experienced a number 
of violent major mergers during the past 6 billion years.
In particular, the most recent major merger occurred at $z\sim 0.2$. 
Most of the major mergers occurred along filaments connecting the Coma 
cluster with other galaxy clusters as shown in Figure \ref{fig_galaxydm}.
This is consistent with the fact that the Coma cluster has two bright 
dominant galaxies, likely the central galaxies of the two massive progenitors 
of the last major merger (at $z\sim 0.2$). Our simulation also reveals an 
ongoing minor merger event on the left side of the cluster 
($\alpha_{\rm J}\sim193.5$ and $\delta_{\rm J}\sim27.5$). This is 
similar to the observed galaxy distribution represented by the 
subgroup, NGC 4839. At $z=0$, the simulated Coma cluster has a mass 
$M_{200}=7.52\times10^{14}\msun$ and a virial radius $r_{200}=1.48\mpc$, corresponding to $\sim$70 arcmins, 
where $M_{200}$ is defined as the mass contained in a spherical region of 
radius $r_{200}$, within which the mean mass density is 200 times the critical 
density. For comparison, the real Coma cluster has a radius of $r_{200}=70$ arcmins, 
according to a scaling relation shown in \cite{Simionescu2013ApJ}.

\begin{figure*}
    \centering
    \includegraphics[scale=0.4]{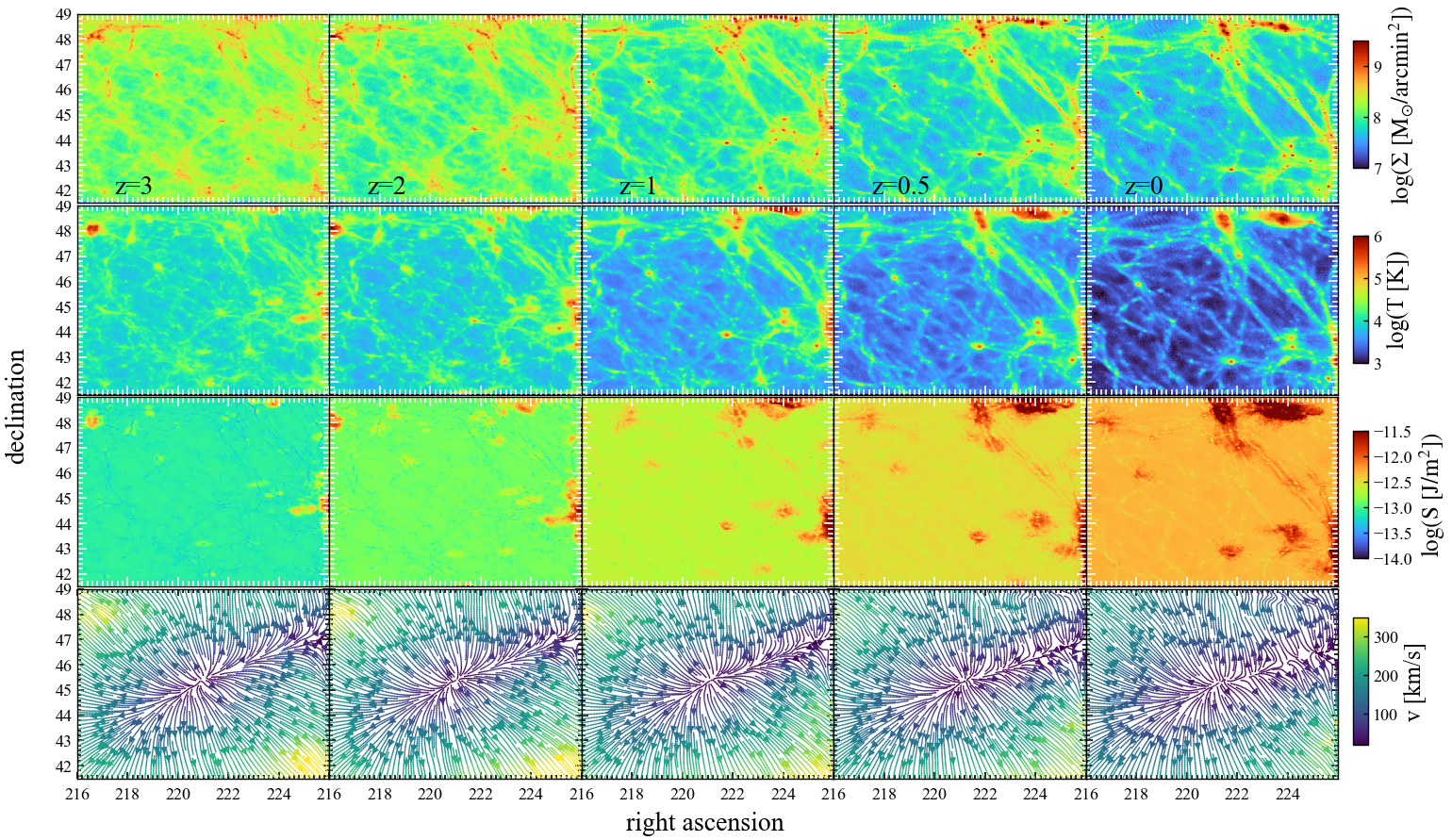}
	\caption{Similar to Figure \ref{fig_comaev} but for the central void region in the LD simulation. Here we present the results from $z=3$ to 0.}\label{fig_voidev}
\end{figure*}

At $z=0.2$ when the major merger starts, the velocity stream lines 
appear to converge at the center of the cluster, as shown in the 
bottom panels of Fig.\,\ref{fig_comaev}. One can see clearly that two 
clumps of gas are colliding with a relative velocity larger than
about $1200\kms$. Such a collision is expected to produce two strong shocks 
propagating in opposite directions \citep[see e.g.][]{Zinger2018MNRAS}.
This is signified by the sharp temperature jump in the inner region
seen in the temperature map at $z=0.1$. The shock propagates outward and, 
at $z=0$, reaches a scale larger than the virial radius of the Coma 
cluster. Meanwhile, a new bow-like shock is produced in the inner region 
(see the density map at $z=0$), indicating that the cluster has not yet 
relaxed to hydrodynamic equilibrium. This result is consistent
with the results obtained from recent X-ray observations 
\citep[e.g.][]{Mirakhor2020MNRAS, Churazov2021A&A}, and 
we will come back to this later in Sections \ref{sec_tsz} and \ref{sec_xray}.

The outgoing shocks can be seen to collide subsequently with the gas 
accretion and heat the gas, as shown in the entropy map. 
This interaction makes the gas stream lines complex. As one can see 
from the $z=0$ velocity map, the infall velocity along the filament 
(from the south-west to the north-east direction) is larger than that 
perpendicular to the filament. This indicates that gas is still 
being accreted along the filament, while, perpendicular to the filament, 
a larger fraction of the gas is driven by outward shocks. 

\subsection{The SDSS Great Wall (GW) simulation}

The gas evolution around massive filaments looks quite different from that around galaxy clusters. In Figure \ref{fig_gwev}, we show the evolution of a massive filament in the GW simulation. Note that this filament is only a small part of the Great Wall.
The whole filamentary structure can already be   
seen at $z>1$. It is consistent with the results presented in \cite{WangH2016ApJ}.
The filament thickens itself by accreting gas from nearby underdense regions.
The velocity streams around the filament have remained at a value 
of $300$ to $400\kms$ since $z=1$ and are roughly perpendicular to the 
filament. The gas accretion onto the filament also leads to accretion shocks. 
As one can see, the gas entropy at the boundary of the 
filamentary structure is higher than that in the inner part
of the filament and the entropy difference increases with decreasing 
redshift, indicating that accretion shocks constantly heat the gas 
surrounding the filament. A similar evolution can also be seen
for the small filaments shown in the figure.  
 
The structure and evolutionary history shown in Fig.\ref{fig_gwev} 
indicate that the diffuse gas in filaments is generally heated by accretion 
shocks, and the heating effect becomes stronger as the structure grows. 
Since the velocity of the gas is lower than that around the Coma cluster, 
the shocks generated are weaker. For example, for the main 
filament shown in the figure, the temperature can reach 
$10^6$-$10^7$ Kelvin. The velocity map shows that the gas flows 
in the filaments are quite turbulent, which may provide an important process 
to mix metal-rich gas generated by galaxies with the IGM.

\subsection{The Low density region (LD) simulation}

In the center of the selected low density region at $z\sim0.05$, there is a large void structure.
Figure \ref{fig_voidev} shows that the structure in the void grows very slowly, 
as expected from the low mass density that drives 
gravitational instability. Gas particles are seen to flow continuously out 
from the void region, making the density decrease with time 
over most of the volume. The gas temperature in the void also 
declines slowly with time. The gas entropy, which is roughly 
homogeneous over the void region, increases with time. 
This is consistent with the fact that gas in low-density regions 
is rarefied by the expansion of the universe and is heated by the UV 
background. As one can see in the density and temperature maps, 
abundant small filamentary structures can be seen since  
$z>2$ and the gas temperature in them has remained in the 
range of $10^{3.5}$ to $10^{4.5}$ Kelvin. However, most of these filaments 
are barely visible in the entropy map, indicating that the formation 
of these filamentary structures is an adiabatic process in an 
ambient medium of similar specific entropy. 
From $z\sim 3$ to the present day, almost all the small filaments have 
an entropy slightly lower than their ambient media, 
indicating that some radiative cooling may be present within them. At $z\sim 0$, 
the entropy of some relatively large filaments appears to be higher than the 
environment, because shock heating starts to become significant.
Clearly, filaments of different sizes have gone through  different heating processes. 

\section{Gas properties and their correlation 
with the cosmic web}
\label{sec_gascor}

\begin{figure*}
    \centering
    \includegraphics[scale=0.6]{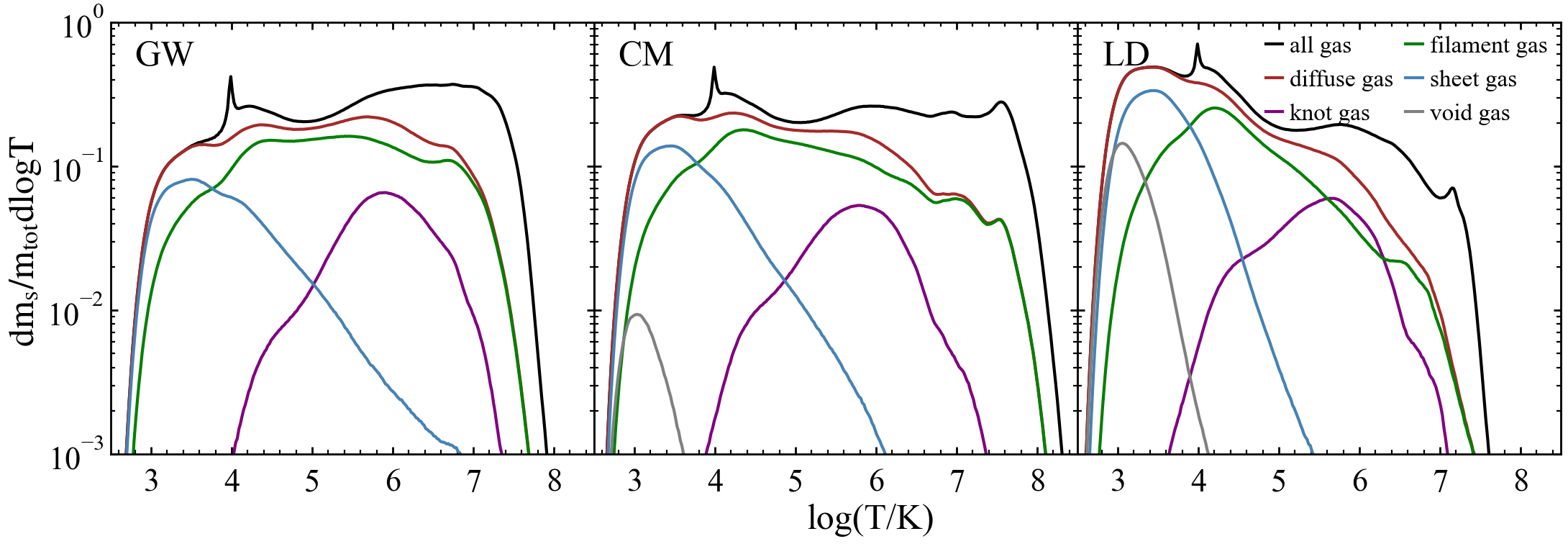}
	\caption{The temperature distribution of gas particles in different types of structures for the GW (left panel), CM (middle panel) and LD (right panel) simulations.
    Black lines are for all gas particles and dark-red lines for all diffuse particles. 
    The other color-coded lines are for gas particles in different components 
    of the cosmic web as indicated. See the text for details of the analysis.
	}\label{fig_Tdis}
\end{figure*}

The above analysis suggests that the gas in different components of the cosmic web
may be heated by different processes. In Figure \ref{fig_Tdis}, we
show the gas temperature distribution in different components of the cosmic web,
as classified using the ``T-Web" method described in 
Section \ref{sec_tweb}. The distribution in a given component is defined as  
${d\rm m_{\rm s}}/{d\rm\log{T}/m_{\rm tot}}$, where 
$d\rm m_{\rm s}$ is the mass of gas particles with a temperature in the range 
of $\rm\log T$ to ${\rm\log T}+d\rm\log{T}$ in the component, 
while $m_{\rm tot}$ is the total gas mass, rather than that in the 
component in question. Note that only gas particles in the HIRs are used 
for the analysis. Thus, the results shown can be used not only to understand  
the temperature distribution within a given type of component, 
but also the contribution from different components. 

\begin{figure}
    \centering
    \includegraphics[scale=0.56]{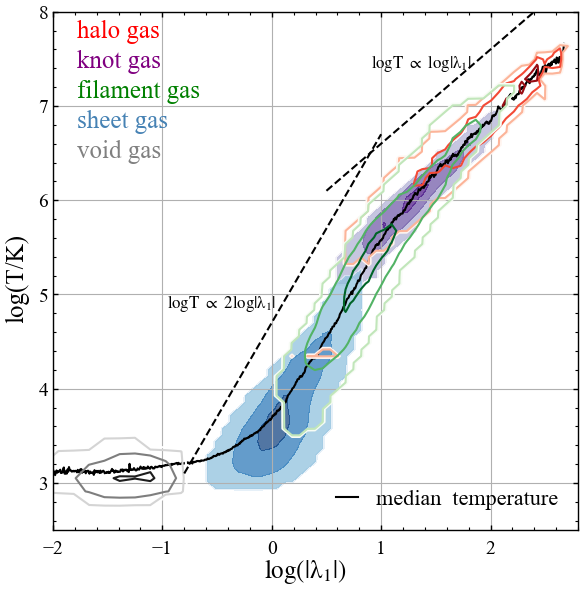}
	\caption{Gas temperature as a function of tidal field strength, $|\lambda_1|$, for different 
	components of the cosmic web (and halos) at $z=0$. Both temperature and tidal field are evaluated on grids. 
	For each component, the contours enclose 10, 45, 80 percents of all the gas mass, 
	respectively. Note that the results for individual components of the 
	cosmic web do not include gas in halos, and that the halo gas shown here does not 
	include gas particles in galaxies. 
	The two dashed lines show two power-law functions with different slopes, as indicated,  
	and the solid line shows the median temperature as a function of the tidal strength 
	for all the gas particles.
	}\label{fig_Tlambda}
\end{figure}

\begin{figure}
    \centering
    \includegraphics[scale=0.56]{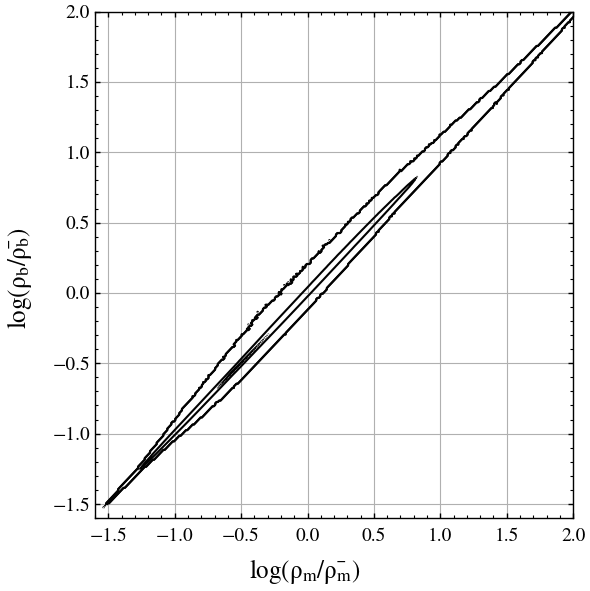}
	\caption{Mass density ($\rho_{\rm m}$) versus baryon density ($\rho_{\rm b}$). 
	Both densities are calculated in cubic cells of $(0.1\mpc)^3$ with a 
	smoothing scale of 0.5$\mpc$, using the GW simulation, 
	and scaled with their mean values, respectively. The 
	contours enclose 10\%, 80\%, 99\% of all cells.
	}\label{fig_dd}
\end{figure}

\begin{figure}
    \centering
    \includegraphics[scale=0.56]{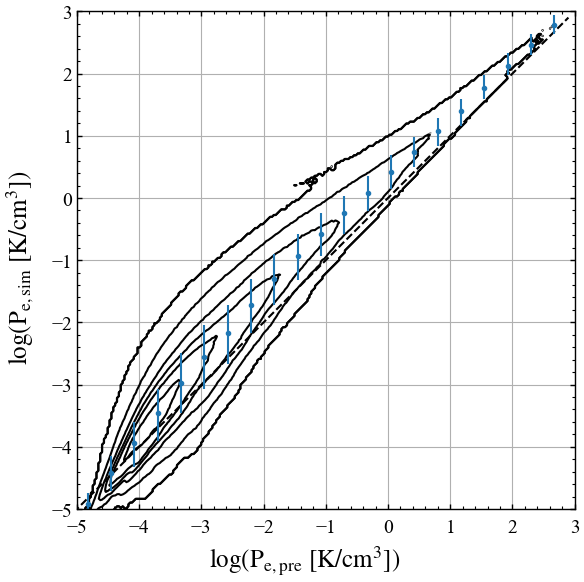}
	\caption{The predicted electron pressure ($P_{\rm e, pre}$) versus the simulated electron 
	pressure ($P_{\rm e, sim}$). The contours enclose, respectively, 
	10\%, 30\%, 50\%, 70\%, 90\%, 99\% of all the cells,
	each being a cube of $(0.1\mpc)^3$. The points show the median 
	$P_{\rm e, sim}$ as a function of $P_{\rm e, pre}$, and error bars show 
	the 16 and 84 percentiles.
	The dashed line corresponds to the one-to-one relation between the predicted and simulated electron pressures. Note that a typical filament with $\rho_{\rm b}=10\bar\rho_{\rm b}$ and $\log(T/K)=5.5$ has a pressure of $0.71\rm K/cm^3$. 
	}\label{fig_pp}
\end{figure}

The black lines in the three panels of Figure \ref{fig_Tdis} show the results for 
all the gas in the three HIRs, respectively.  
The gas temperature distributions in the three regions are clearly different. 
In the GW simulation, there is a significant bump around $\log(T/K)\sim 7$, 
which is absent in the other two regions. This is produced by the existence of 
the massive filamentary structure in the GW. In the CM simulation, the gas temperature 
can reach as high as $10^8K$ because of the presence of the massive Coma cluster. 
Finally, the LD simulation is dominated by low-temperature gas, as expected.

The temperature distribution for gas outside halos (diffuse gas) is also shown 
in each of the three panels. As one can see, the spike at $\log(T/K)\sim4$ 
for all gas particles disappears, indicating that this feature is dominated by 
cold CGM in halos. The diffuse gas dominates at 
the low-temperature end and becomes less important as the temperature increases. 
At $\log(T/K)=5$, more than 80 percents of the gas is in the diffuse component, and 
the fraction decreases to about 20 percents at $\log(T/K)=7$. About half
of the gas with $5<\log\rm (T/K)<7$ (the WHIM) is located outside halos, 
quite independent of the large-scale environment. This result is
consistent with that obtained before \citep[e.g.][]{Cen1999ApJ, Dave2001ApJ, Martizzi2019MNRAS}.

In Figure \ref{fig_Tdis} we also plot the temperature distribution of  
the diffuse gas separated into the four different components of the cosmic web. 
In most cases, the gas in filaments dominates the diffuse gas   
at $\log\rm(T/K)>4$. This indicates that most of the warm-hot diffuse gas 
resides in filaments. At $\log\rm(T/K)<4$, sheet gas contributes the most. 
Void gas has a very narrow distribution peaked at the lowest temperature, 
and its contribution is important only in the LD simulation. 
Our results also show that the diffuse gas with the highest temperature 
is mostly associated with filaments rather than knots, which is different 
from the expectation that high temperature gas is associated with 
the outskirts of massive clusters.  We note that in detail the results
depend on the smoothing scale $R_{\rm s}$ used for the analysis.  
If a larger value of $R_{\rm s}$ were used, hot gas in the interior 
of halos could affect the estimate of gas properties 
in the outskirts of halos. For example, if we adopt $R_{\rm s}=2\mpc$,
instead of $R_{\rm s}=0.5\mpc$ as used above, the diffuse gas associated 
with the knot component becomes hotter than the filament gas.
The small smoothing scale, $R_{\rm s}=0.5\mpc$, used here minimizes the contamination by halo gas in our analysis of the 
diffuse components. 

Given the differences in the gas temperature between different components of the cosmic web as shown in Figure \ref{fig_Tdis}, it is interesting to directly check the correlation between the temperature and the local tidal field that was used to decompose the cosmic web.
In Figure \ref{fig_Tlambda}, we show the gas temperature as a function
of $|\lambda_1|$ for halo gas and diffuse gas in the four components of the cosmic web.

As an example, here we only show the results for the GW simulation.
The results for the other two simulations are similar, although 
there are differences in their temperature distributions (Figure \ref{fig_Tdis}).

As one can see, there is a strong and tight correlation between the gas temperature 
and the tidal field strength represented by $|\lambda_1|$. 
The correlation can be separated into three distinct 
components. For illustration, the two dashed lines
show $\log T\propto 2\log\lambda_1$ and $\log T\propto \log\lambda_1$, 
respectively. As one can see, the halo gas and the diffuse knot gas 
are both dominated by hot gas with $T\gs 10^6{\rm K}$ and both follow 
the relation $\log T\propto \log\lambda_1$, suggesting that
both are heated by similar processes. The gas in sheets is warm, with a  
temperature between $10^{3}$ - $10^5{\rm K}$, follows
$\log T\propto 2\log\lambda_1$, and has almost no overlap with 
the halo and knot components. The gas in filaments has a more complex behavior. 
At high temperature, it follows the relation of the hot gas, 
while at low temperature, it obeys the correlation of the warm (sheet) component. 
The dividing point is around $\log T\sim5.4$ where $\log|\lambda_1|\sim 1$.
Finally, void gas, which is relatively cold, has a temperature 
quite independent of the tidal field strength.

The tight correlation shown above suggests that the local tidal field strength 
is a good indicator of the gas temperature before radiative cooling becomes important. 
It may thus be possible to reconstruct the three-dimensional gas pressure distribution 
from the reconstructed total mass density field. To do this, we need to check how 
well the gas density correlates with the total mass density. 
Figure \ref{fig_dd} shows the gas density versus mass density 
obtained by using a smoothing scale of $R_{\rm s}=0.5\mpc$. 
The two densities are tightly and linearly correlated. 
To predict the pressure at a given location, we first use the value of 
$\lambda_1$, again smoothed with $R_{\rm s}=0.5\mpc$, to predict the 
gas temperature using the median relation shown in Figure \ref{fig_Tlambda}. 
We obtain the gas density, $\rho_{\rm g}$, from the total mass density, 
$\rho_{\rm m}$, using a simple relation $\rho_{\rm g}=f_{\rm b}\rho_{\rm m}$, 
where $f_{\rm b}$ is the universal baryon fraction. The electron pressure 
at the grid point is obtained using $P_{\rm e, pre}=kn_{\rm e}T$, assuming fully 
ionized gas and primordial metallicity. In Figure \ref{fig_pp}, we show the comparison of the
electron pressure predicted in this way with that obtained directly from the simulation. 
As one can see, the predicted pressure is tightly correlated with the simulated value, 
with a typical scatter of less than 0.5 dex, and the scatter increases with decreasing pressure. 
The method recovers the simulation results with only small bias (up to 0.5 dex)
that also increases with decreasing pressure. This dependence of 
the bias and scatter on pressure is caused by the weak correlation 
between tidal field strength and temperature at the low end of $\vert \lambda_1\vert$.
For a typical filament with $\rho_{\rm g}=10\bar\rho_{\rm b}$ and 
$\log(T/K)=5.5$, the electron pressure is about $0.71\rm K/cm^3$, and 
the scatter between the reconstruction and simulation is expected to be less than 0.25 dex.
Thus, our pressure reconstruction is reliable at least for filaments and large sheets.
As discussed in Section \ref{sec_tsz}, this method could be applied to 
constrained, pure dark matter simulations to reconstruct the temperature, 
gas density and pressure fields in regions where accurate reconstructions 
of the dark matter density field are available. 

\section{Connections to observational data}\label{sec_obs}

In this section, we use our constrained simulations to generate a number of 
observable quantities, such as the expected thermal Sunyaev-Zel’dovich (SZ) effect, 
X-ray emission, and HI and OVI absorption systems.  
These results can be compared with observational data statistically or 
for individual objects, taking full advantage of our constrained simulations.

\subsection{The Thermal SZ effect}\label{sec_tsz}

\begin{figure*}
    \centering
    \includegraphics[scale=0.6]{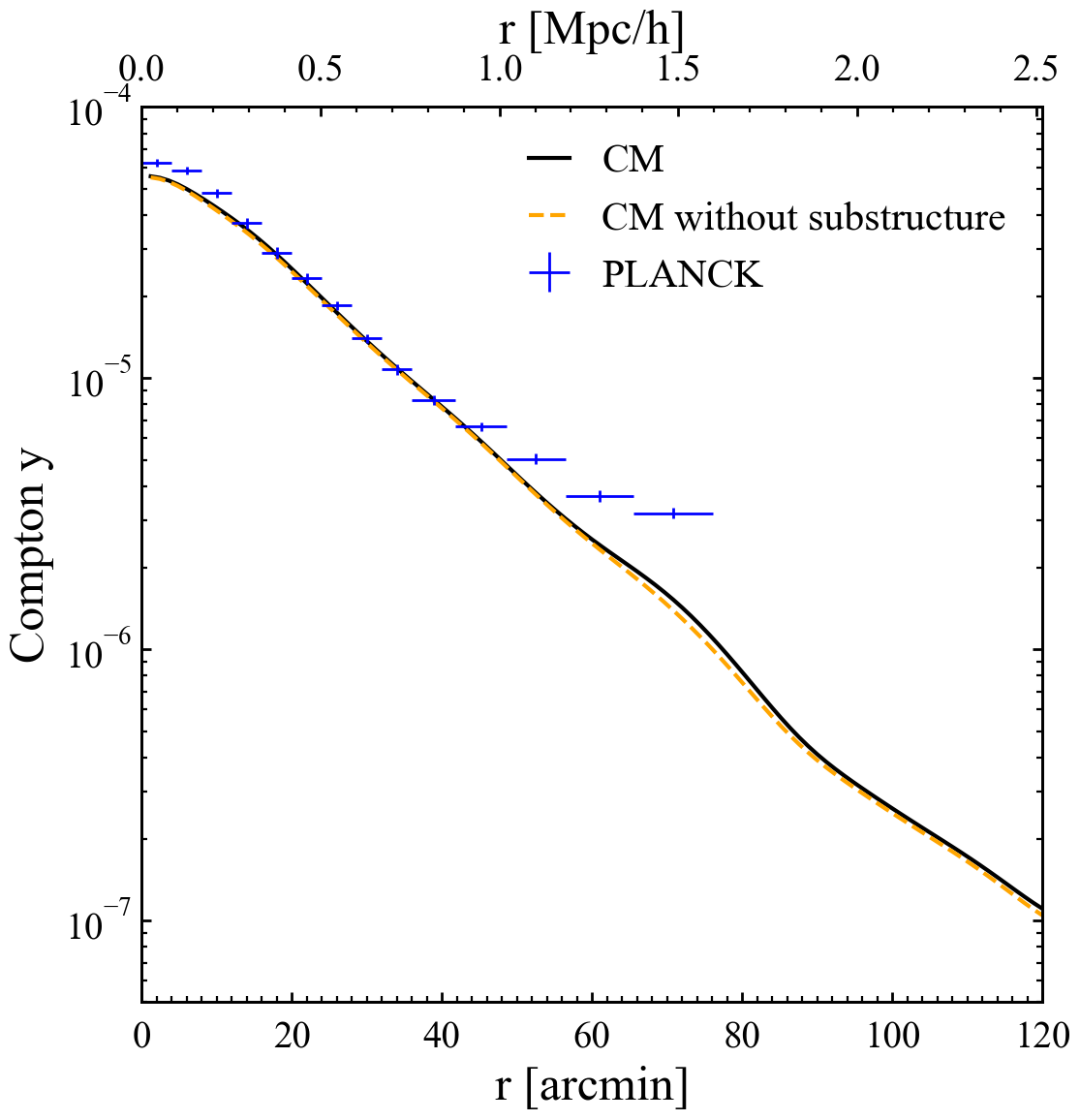}\includegraphics[scale=0.6]{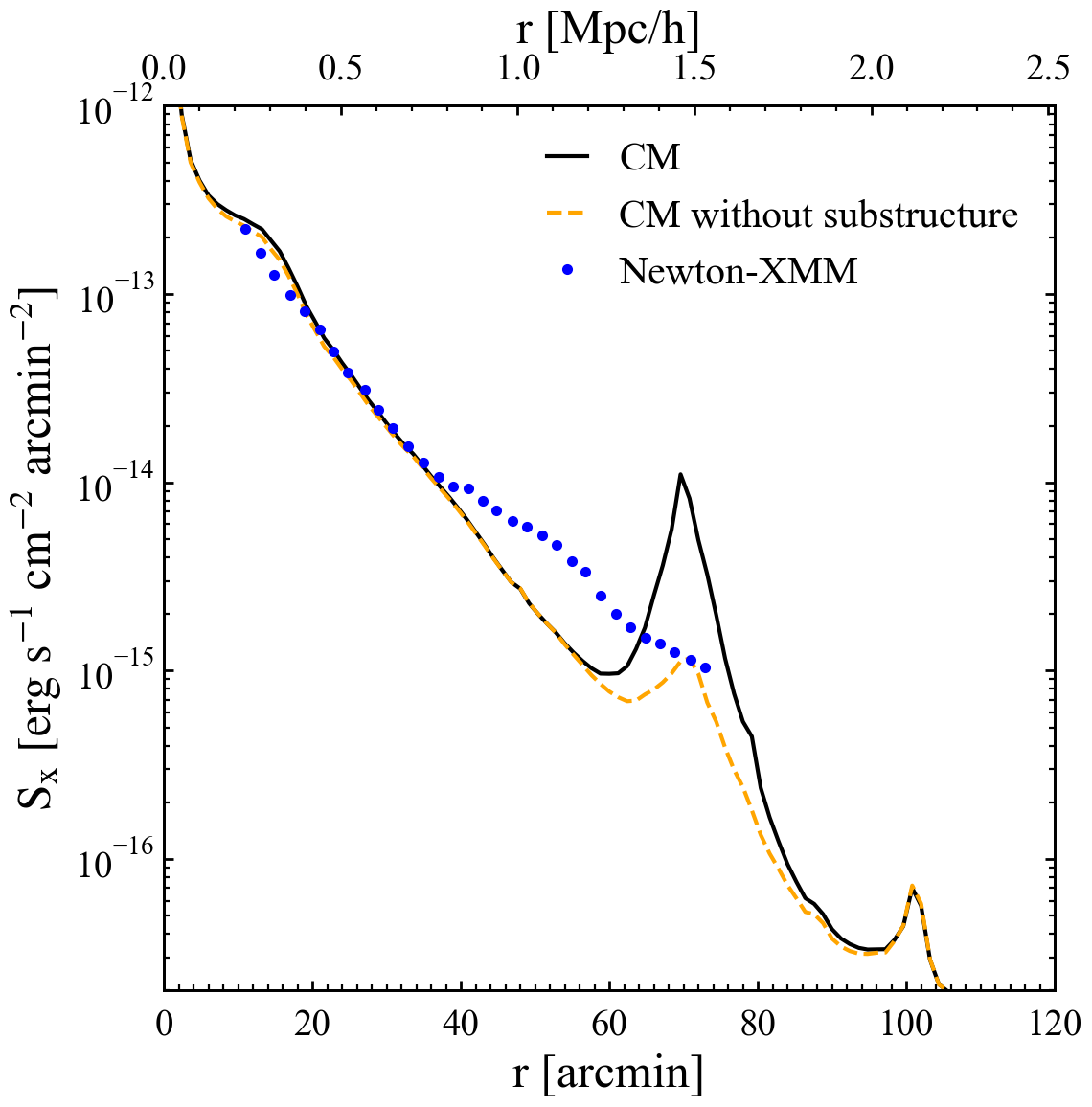}
	\caption{Left panel: The Compton $y$-profile (azimuthally-averaged) for the 
	Coma cluster. The data points, taken from \cite[][private communication]{Mirakhor2020MNRAS}, 
	are the measurements from Planck and the lines show the results from our CM simulation. 
	The solid black line shows the result averaged over the whole Coma cluster, while the 
	dashed curve shows the result after the removal of the most significant 
	substructure (see Figure \ref{Fig_comaxrayYmap}). 
	To compare with the observational data, the simulation result is smoothed with a 
	${\rm FWHM}=10~\rm arcmin$. Right panel: The surface brightness profile 
	of X-rays in the band of 0.7-7.0~keV for the Coma cluster. 
	The data points are based on Newton-XMM observations \citep[][private communication]{Mirakhor2020MNRAS}, and the two lines show results with and without the substructure removed, 
	respectively.
	}\label{fig_comaxyprof}
\end{figure*}

The thermal Sunyaev Zel’dovich (tSZ) effect provides an important avenue to investigate 
the ICM and IGM \citep{SZ1972CoASP}. 
The strength of the tSZ effect in a given direction is described by the dimensionless Compton parameter, 
$y$, defined as 
\begin{equation}
    y=\frac{\sigma_{\rm T}}{m_{\rm e}c^2}\int P(l)dl\\,\label{eq_y}
\end{equation}
where $m_{\rm e}$ is the electron mass, $\sigma_{\rm T}$ is the Thomson 
scatter cross-section, $c$ is the speed of light, $P$ is the electron pressure, and 
the integration is along a given line of sight. 
For each simulation, we determine the ranges of $\alpha_{\rm J}$ 
and $\delta_{\rm J}$, over which the tSZ effect is to be modelled, 
and divide the sky into two-dimensional pixels of a given angular size. 
We thus obtain a $y$-parameter map sampled on all the pixels in the sky coverage.
This map is also used to study the profiles around 
individual objects, such as the Coma galaxy cluster. 

\begin{figure}
    \centering
    \includegraphics[scale=0.65]{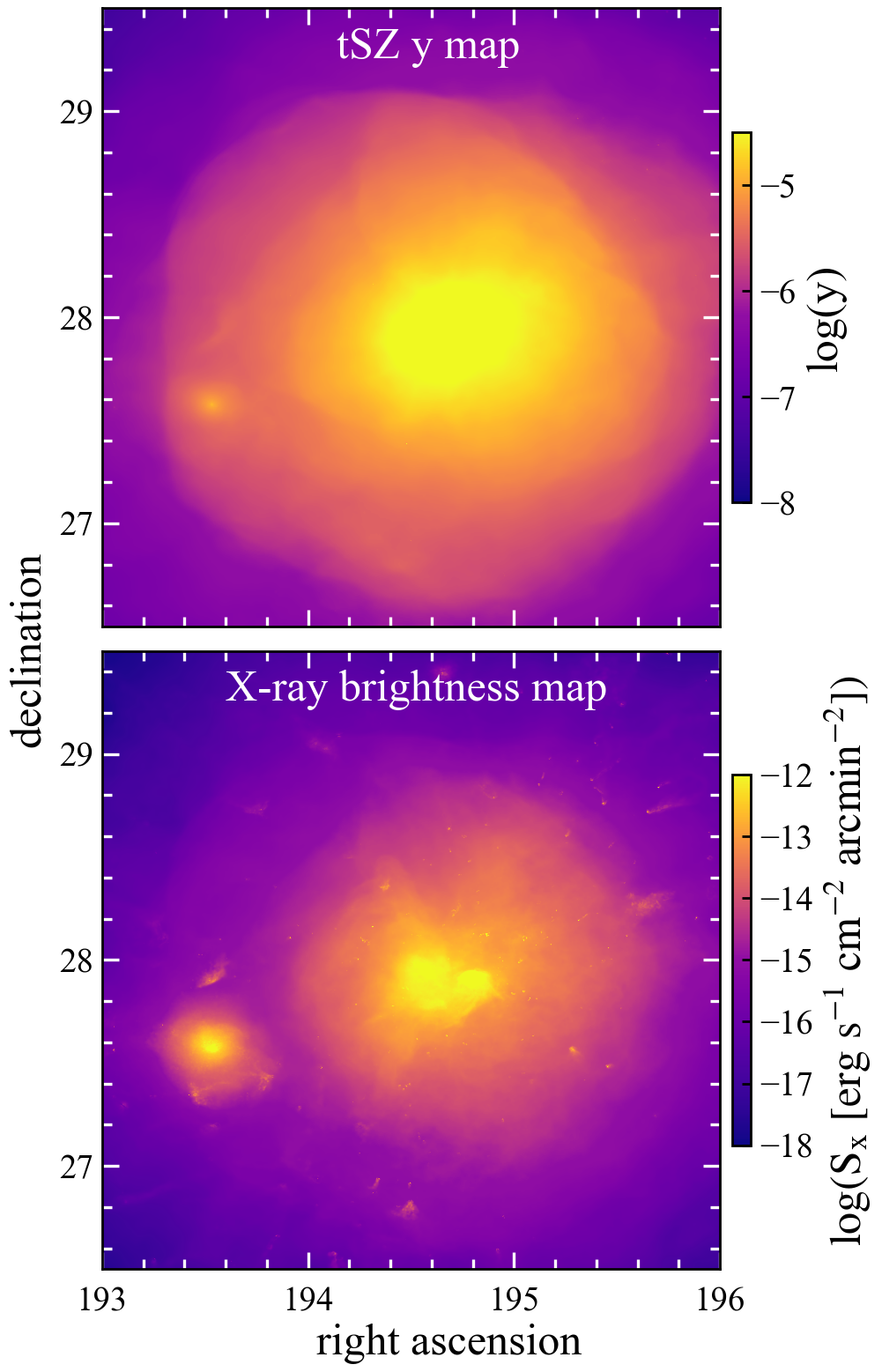}
	\caption{Maps of the $y$-parameter (upper panel) and 
	X-ray (0.4-2 KeV band) brightness (lower panel) around the Coma cluster in the CM simulation.}\label{Fig_comaxrayYmap}
\end{figure}

We adopt two methods to construct $y$-maps. In the first one, 
we divide the volume covered by a 2-d pixel into many small cells. The cell 
sizes are chosen to be smaller than the gravitational softening length of $1.8\kpc$. 
We estimate the number density and temperature of each cell from nearby gas particles 
using the SPH smoothing kernel. We then integrate along the line
of sight (Eq. \ref{eq_y}) to obtain the $y$-parameter of each pixel.
The results presented in Figure \ref{fig_comaxyprof} and \ref{Fig_comaxrayYmap} are 
obtained with this method (hereafter SphM). The SphM method is very time consuming 
and also unnecessary for some of our analyses. We therefore adopt a second, faster 
method for some of our presentations. Specifically, for a 2-d pixel with a 
solid angle of $\Delta\Omega$ covering $M$ gas particles, we approximate Eq. \ref{eq_y}
with 
\begin{equation}
    y=\frac{\sigma_{\rm T}k_{\rm B}}{m_{\rm e}c^2\Delta\Omega}
    \sum_i^{M}\frac{N_{{\rm e},i}T_i}{r_i^2},\label{eq_yp}
\end{equation}
where $N_{{\rm e},i}$ and $T_i$ are, respectively, the number of free electrons 
and gas temperature represented by particle $i$, $r_i$ is the distance of the particle, 
and $k_{\rm B}$ is the Boltzmann constant. This approach (hereafter ParM) is valid when the 
pixel size is much larger than the local SPH kernel radius. 
As shown in the Appendix, the ParM method is less accurate than the SphM, 
especially in low density regions where the SPH kernel is large, but is 
sufficient for the purpose of this paper.

The left panel of Figure \ref{fig_comaxyprof} shows the radial profile of the 
$y$-parameter of the Coma cluster, obtained from the simulated $y$-map using SphM.
The result is extended to an angular radius of 120 arcmins, which is about 1.7 times
the virial radius of the cluster. For comparison, we also show the observed 
$y$-parameter profile for the Coma cluster 
obtained in \cite{Mirakhor2020MNRAS} from the Planck $y$-map
\citep[e.g.][]{PlanckX2013A&A}. Since the effective spatial resolution 
of the Planck $y$-map is about $10~\rm arcmin~{\rm FWHM}$ (full width at 
half-maximum), we smooth the simulation result with the same FWHM.
As one can see, the simulated profile 
matches the observed one well at $\theta<50$ arcmins, corresponding 
to about 0.7 times the virial radius, but is lower around the virial radius. 
Recently, \cite{Anbajagane2021arXiv} analyzed the SZ data of several hundred 
clusters and compared the results with simulations\citep{Cui2018MNRAS}. They found that the average 
profile of the simulated clusters is also lower than the observation
outside the virial radius. They suggested that the discrepancy might owe to the simulations not properly handling 
non-equilibrium effects, which might be significant because of the 
presence of accretion shocks and the low gas density in the outskirts of clusters.  
Another possibility is that our constrained simulations do not model accurately
potentially important processes in the central part of the Coma cluster, 
such as the radio mode of AGN feedback. We will investigate these issues 
in future work. 

Figure \ref{Fig_comaxrayYmap} shows the $y$-parameter map 
around the simulated Coma cluster. One can see a large, vertical edge
located  at $\alpha_{\rm J}\sim193.3$, which connects to a
large edge extending from $(\alpha_{\rm J}\sim193.3, \delta_{\rm J}\sim 27)$ to 
$(194.4, 26.6)$. This feature is outside the virial radius and can 
also be seen in the temperature and entropy maps shown in Figure \ref{fig_comaev}. 
In the $y$-map of the Coma cluster obtained from PLANCK, one can also see 
a rapid decline of $y$ at a similar location \citep[see Figure 3 in][]{PlanckX2013A&A},  
and a diffuse radio relic is also observed there 
\citep[see Figure 1 in][]{Brown2011MNRAS}. An inspection of the simulation 
snapshots shown in Figure \ref{fig_comaev} reveals that the shock induced 
by the merger event at $z\sim0.2$ produces this feature.  
In the $y$-profile of the simulated Coma shown in
Figure \ref{fig_comaxyprof}, one can also see a significant drop at $r\sim 80~{\rm arcmin}$, which 
roughly corresponds to the sharp edges seen in the $y$-map. Note that there is a substructure
at a similar radius to the left of the main 
cluster, which might also contribute to the drop in the 
$y$ profile. The dashed curve in the left panel of Figure \ref{fig_comaxyprof}
shows the result excluding the region around the substructure. 
Clearly, the drop seen is not affected significantly by this substructure.  
All these demonstrate that the region around the Coma cluster provides an 
ideal avenue to detect and study the diffuse gas produced by accretion and merger driven 
shocks.

\begin{figure*}
    \centering
    \includegraphics[scale=0.8]{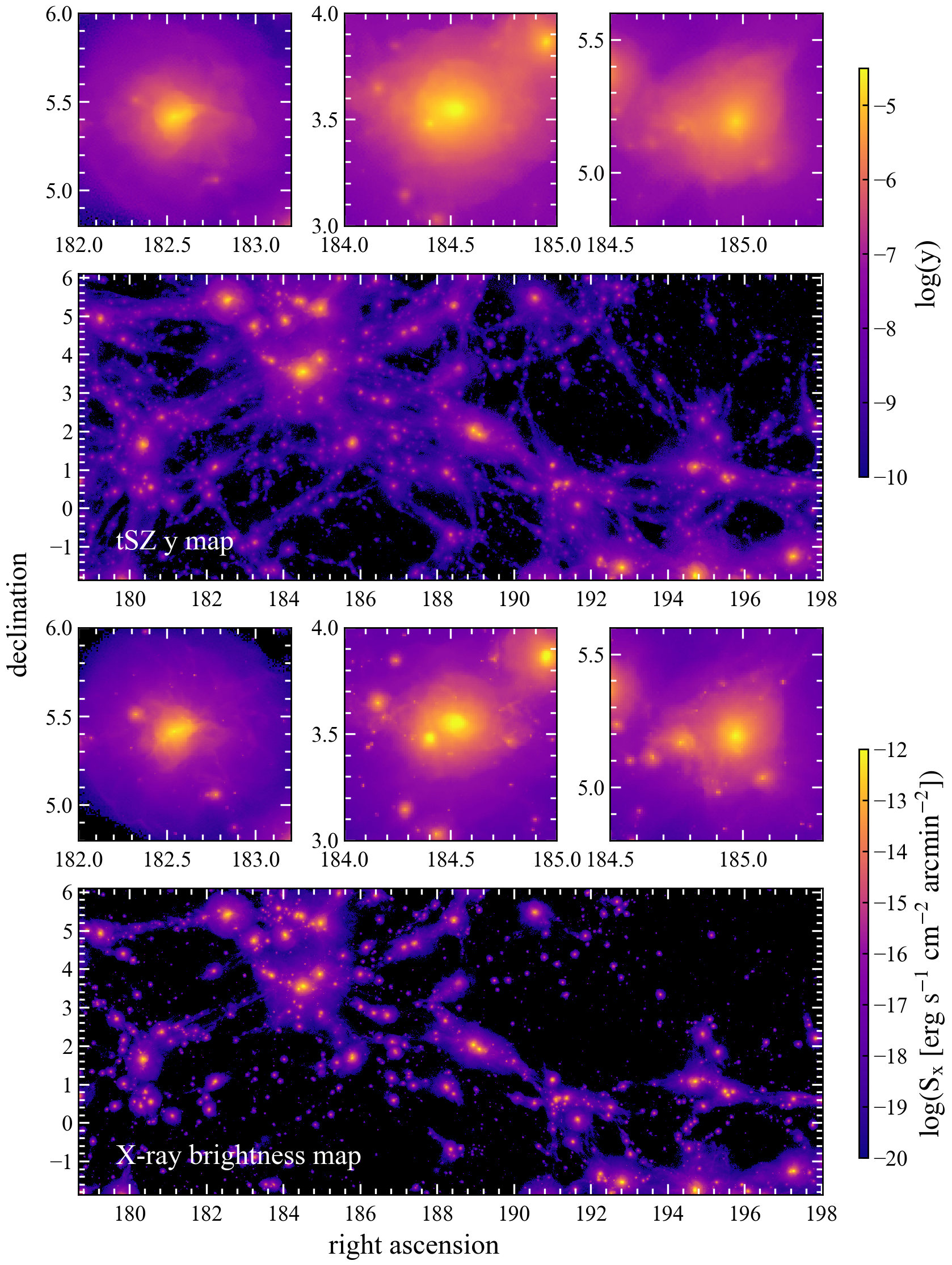}
	\caption{Maps of the $y$-parameter (the upper two rows) and X-ray (0.3-2 KeV band) brightness 
	(the lower two rows) in the GW simulation. The first and third rows show the 
	results for the three massive galaxy clusters in the super-cluster complex located at 
	$\alpha_{\rm J}\sim184$ and $\delta_{\rm J}\sim4.5$.
	}\label{Fig_gwxrayYmap}
\end{figure*}

Recently, attempts have been made to detect the SZ effect 
produced by filaments. However, only a signal between two merging 
massive clusters was claimed to be detected\citep[e.g.][]{Planck2013VIII, deGraaff2019}. 
It is because the gas temperature and density in 
filaments is usually much lower than those in clusters and a merging 
system can boost them. The SDSS great wall region contains many 
small and massive filamentary structures.
Figure \ref{Fig_gwxrayYmap} shows the Compton $y$-map in the GW simulation.
This map is obtained with the ParM method using a pixel size of 0.5 arcmin.
As shown in the Appendix, this method is reliable at $\log y>-8.5$ for such a pixel size.
As one can see, the logarithm of the $y$-parameter in filaments is 
usually between -9 and -8, much lower than the detection
limit of current observations. The super-cluster complex located
at $\alpha_{\rm J}\sim184$ and $\delta_{\rm J}\sim4.5$   
consists of three clusters with halo masses above $10^{14}\msun$ 
in a volume of $(10\mpc)^3$ and has a mean density about $17.2$ times the cosmic mean density. 
A large fraction of this particular region is covered by pixels with 
$\log y>-8$, which may still be too weak to be detected directly by
current observations. However, it may be possible 
to stack many pixels to have an average detection. Such a technique has 
already been applied successfully to detect the SZ effect produced by the  
CGM and the IGM \citep[e.g.][]{Lim2018ApJ,Lim2018MNRAS}.

Inspecting a series of snapshots of this region 
reveals a large number of violent merger events. 
These mergers are expected 
to produce strong shocks that could propagate outwards and heat the entire region.
It may explain why this region has a larger $y$ than other regions.
In Figure \ref{Fig_gwxrayYmap}, we also show $y$-maps around the three massive clusters in the super-cluster complex. As one can see,  
there are many interfaces with sharp discontinuities around them. 
Some show sharp discontinuities at both small and large scales, reflecting 
shocks induced by ongoing interactions in the recent merger histories of 
these systems. Our constrained simulations will provide a unique data set to 
select interesting structures, such as filaments, for a stacking 
analysis, particularly when surveys of high sensitivity and 
resolution, such as CMB-S4\citep{Abazajian2016CMBS4}, become available.

Our constrained simulations can also be used for cross-correlation investigations.
For example, one can cross-correlate the simulated $y$ map with observational data
to study how well the model predictions match observations. 
Owing to the weakness of the SZ effect,  
such cross-correlation analyses usually require large hydro-dynamical simulations 
with a large sky coverage, which are time consuming to run. However, 
as shown in Section \ref{sec_gascor}, the local tidal strength $\lambda_1$
and the dark matter density $\rho_{\rm m}$ are tightly correlated with the  
gas temperature $T$ and density $\rho_{\rm gas}$, respectively. 
Thus, one can reconstruct the gas pressure field in the cosmic web. 
We can apply this method to the whole ELUCID $N$-body simulation 
in the SDSS region \citep{WangH2016ApJ}, or any other constrained 
cosmological $N$-body simulations, to predict the spatial distribution 
of electron pressure in the SDSS region. Assuming a simple relation between 
electron pressure and mass density, \cite{Lim2018MNRAS} 
obtained a model pressure map based on the ELUCID density field in 
the SDSS volume, and cross-correlated the model pressure
with the Planck $y$-map to constrain the pressure - mass density relation
over a large range of mass density. In particular they found that, 
at a given density, the $y$-parameter increases with the 
value of $\vert \lambda_1\vert$, as expected from its correlation with the 
gas temperature. The tight $\lambda_1-\rm T$ relation 
found above suggests that such an analysis can be made much more powerful 
by incorporating the reconstructed local tidal strength.  
As shown in Figure \ref{fig_pp}, the typical scatter of the 
reconstructed pressure is about 0.25 dex over the density range 
relevant for filaments and sheets. This will provide a much improved 
criterion to select targets for stacking analyses.

\subsection{X-ray emission}\label{sec_xray}

To calculate the X-ray surface brightness in a given band of $[E_1, E_2]$, we divide the 
sky of interest into many small pixels, each with a solid angle of $\Delta\Omega$. 
For each pixel, we calculate the surface brightness by summing up the contribution of 
all gas elements (cells in SphM method or particles in ParM method) in the pixel,
\begin{align}
\begin{split}
S_x(E_1, E_2)
&=\frac{1}{\Delta\Omega}\sum_i\frac{1}{4\pi D^2_L(z_i)} 
\\
&\int_{E_1(1+z_i)}^{E_2(1+z_i)}\Lambda(E,T_i,Z_i)n_{{\rm H},i}N_{{\rm e},i}dE,
\end{split}\label{eq_xray}
\end{align}
where the emissivity, $\Lambda$, is taken from look-up tables provided by 
the AtomDB \citep[][version 3.0.9]{Foster2020Atoms}\footnote{http://www.atomdb.org/index.php};
$n_{{\rm H},i}$, $N_{{\rm e},i}$, $T_i$ and $Z_i$ are the number density of hydrogen ions, 
the number of free electrons, the temperature and metallicity of the gas element $i$; $z_i$ and $D_L(z_i)$ are the cosmological 
redshift and luminosity distance of the gas element, respectively, 
calculated according to the distance of the element to Earth. 
The emissivity, $\Lambda$, provides both continuum and line emission under 
the assumption of collisional equilibrium, and all emission lines available in the 
chosen X-ray band are taken into account. Note that we exclude the contribution 
of gas elements within galaxies to the X-ray emission.

As in the last subsection, we use SphM to calculate the X-ray emission from the Coma cluster, and ParM for the GW simulation. 
A comparison between the two methods for X-rays is presented in the Appendix.
Figure \ref{fig_comaxyprof} shows the X-ray surface brightness 
profile of the simulated Coma cluster in the band of $[0.7,7]$ keV. 
For comparison, we also show the observational data from XMM-Newton 
\citep[][private communication]{Mirakhor2020MNRAS}. Similar to the $y$-parameter,
our simulation is in good agreement with the observational data
in the inner region from 10 to 40 arcmin. 
In the most inner region, the simulation result shows a sharp rise. 
As shown in \cite{Huang2019MNRAS}, simulations 
using the PESPH technique may produce more dense gas 
in the central region of halos than the 
traditional SPH technique \citep[e.g.][]{Springel2005MNRAS}. 
The small peaks on the X-ray profiles may also be caused by this artificial dense gas.
In addition, our simulation does not include AGN feedback, 
and it is possible that AGN feedback can 
reduce the amount of high-density gas near the center. 
At $r>40$ arcmins, the simulation prediction is systematically 
lower than the observation. The peak 
at $r\sim 70$ arcmin in the simulation result is caused by the 
substructure discussed in the above subsection, and is significantly 
suppressed after the removal of the substructure (see the dashed curve).  

Figure \ref{Fig_comaxrayYmap} shows the X-ray brightness map around the 
simulated Coma cluster. 
To see the inner structure more clearly 
and to compare with the eROSITA observations \citep{Churazov2021A&A},
the brightness profile is obtained in the band of $[0.4, 2.0]$KeV. 
One can see a bow-like sharp structure, more than one degree long, to the left
of the cluster. There is another significant feature that is 
luminous and also has sharp edges at the center of the cluster.
Interestingly, both features are seen in the eROSITA X-ray image
of \citet{Churazov2021A&A}, who suggested that the  
feature to the left is likely produced by a shock
while the one at the center is a contact discontinuity. 
Inspecting the formation history in our simulation indicates that the 
two features are actually generated by a merger that occurred more 
recently at $z\sim 0.2$.

In general, it is quite challenging to use X-ray emission to probe 
diffuse gas in the cosmic web, although attempts have been made to detect 
X-ray filaments associated with merging clusters 
\citep[e.g.][]{Sugawara2017PASJ, Reiprich2021A&A}. Figure \ref{Fig_gwxrayYmap} 
shows the predicted X-ray brightness map in the GW region, calculated in the 
band of [0.3, 2.0]KeV. The pixel size of this map is 0.5 arcmin. 
As shown in the Appendix, at such a pixel size the ParM method 
is reliable at $\log S_{\rm x}>-19$.
The X-ray brightness drops very quickly as the distance to
massive halos increases, because both the gas density and temperature decrease rapidly
with the distance. The super-cluster around $\alpha_{\rm J}\sim184$ and 
$\delta_{\rm J}\sim4.5$ might be a potential place to study diffuse gas 
in X-ray, as several relatively massive halos reside there. 
As a demonstration, we show the X-ray maps of the three most massive clusters in the 
region. As one can see, shock edges are also present around these clusters. 
The Cosmic Web Explorer, a proposed X-ray observatory, expected to 
reach $S_{\rm x}=10^{-18} {\rm erg s^{-1} cm^{-2} arcmin^{-2}}$  in the band of 
[0.3, 2.0]KeV\citep{Simionescu2021ExA}, would make the X-ray emission from the diffuse 
gas predicted here easily detectable.

The good agreement between our simulation results and observational 
data in SZ and X-ray profiles, as well as in the locations of 
shock fronts and discontinuity features, in Coma cluster suggests that our simulation 
can reproduce the merging and heating histories of the Coma 
cluster reasonably well. Our simulations can thus provide a 
powerful avenue to interpret observational data. 
Our simulation results also have implications for AGN feedback. 
The fact that the predicted gas density and temperature profiles
match the observed SZ and X-ray profiles at 10 to 40 arcmins
suggests AGN feedback should not affect the gas significantly on 
these scales, because our current simulations do not include AGN feedback. 
Observations of both SZ effects and X-ray emissions of galaxy groups suggest that
AGN feedback may be able to push a significant fraction of baryonic gas out 
of their halos and heat the IGM in the cosmic web
\citep[e.g.][]{SunM2009, Lovisari2015, Amodeo2021, Schaan2021}. 
It is thus possible that the AGN feedback can dramatically change the  
X-ray and SZ predictions in regions of intermediate gas densities. 
The predicted X-ray and SZ signals in filaments shown in 
Figure \ref{Fig_gwxrayYmap} may thus be sensitive to the implementation of 
AGN feedback. Clearly, a detailed comparison between simulations 
of different AGN feedback and observational data are expected
to provide important constraints on the underlying physical processes
\citep[see, e.g.][]{Lim2018MNRAS}.
We will come back to this in a forthcoming paper. 

\subsection{Quasar absorption systems}\label{sec_abs}

\begin{figure*}
    \centering
    \includegraphics[scale=0.9]{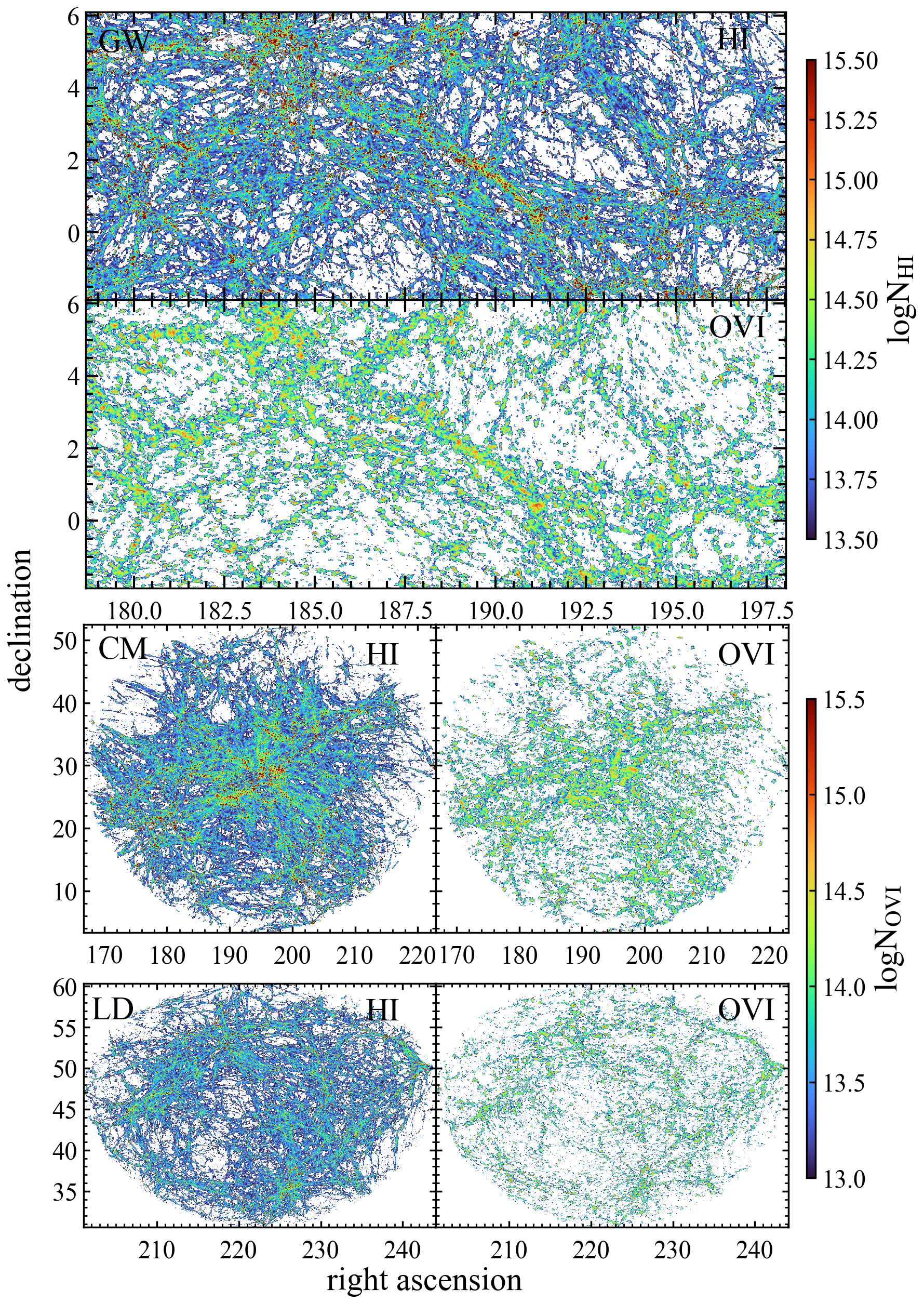}
	\caption{The HI and OVI column density maps obtained from the three regions as labelled. }\label{fig_absmap}
\end{figure*}

Figure \ref{fig_absmap} shows HI and OVI column density maps in the three sky regions.
To obtain these column density maps, we need the ionization states of gas particles.
We use look-up tables from SPECEXBIN \citep{Oppenheimer2006MNRAS,Ford2013MNRAS} 
to find the ionization fraction for any gas particle in our simulations.
These tables provide ionization fractions for various ions as functions
of gas density, temperature and redshift, obtained from CLOUDY \citep[][version 08]{Ferland1998PASP} 
assuming an optically thin slab of gas and Collisional and photoionization 
equilibrium with a uniform \cite{HM12} background (hereafter HM12).
For each sky region, we first select a large set of lines of sight  that have a uniform 
distribution on the sky. We then divide each line of sight into 
many tiny intervals, each with a length equal to the gravitational softening 
length of $1.8\kpc$. For each interval, we estimate the number 
density of the ion species in question from nearby gas particles
using the SPH smoothing kernel.

Then we calculate the column density of an absorption line by 
integrating the density along the line of sight within the high resolution region. 
As shown in \cite{Huang2019MNRAS}, assuming the
HM12 UV background leads to a systematic overestimate of the 
HI column density. Following \cite{Dave2010}, \cite{Huang2019MNRAS} adjusted 
the HI optical depth by matching the evolution of Ly$\alpha$ flux decrements to 
observations and obtained a correction factor of 0.31. We thus multiply all HI column 
densities by the same factor. It should be pointed out, however, that the column 
density calculated in this way cannot be compared directly with the observed 
column density of individual absorption lines, as our analysis does not decompose 
the absorption into individual lines.

As one can see, high-$N_{\rm HI}$ gas, e.g. $\log N_{\rm HI}>15$, has a very compact
distribution, while low-$N_{\rm HI}$ gas resides in filamentary structures. 
This is expected for low ionization energy ions, such as HI, which 
can be more easily destroyed by photoionization in gas of lower density.    
Almost all OVI absorption systems with $\log N_{\rm OVI}>13$ are produced  
by diffuse gas in filaments. These results indicate that low $N_{\rm HI}$ Ly$\alpha$ 
absorption and most of the OVI absorption with $\log N_{\rm OVI}>13$ 
are good tracers of the diffuse gas in filaments \citep[see below and][]{Bradley2022}. 
Compared to the OVI maps, the HI maps contain many smaller filamentary 
structures, indicating that low column-density Ly$\alpha$ absorption can 
be used to probe smaller filaments in the cosmic web, while OVI absorption 
mainly traces larger filaments. Halos residing in
these thin filaments are usually very small, and thus have very 
low star formation efficiency \citep[see e.g.][]{Zhang2021} and low  
metal abundances to produce significant OVI absorption. 
To verify this, we generated metallicity maps and found that massive 
filaments are well represented, while small filaments disappear, 
in the metallicity maps. 

One can also see that the HI maps contain many more 
compact clumps of high column density than the OVI maps. 
As shown in \cite{Nelson2018MNRAS}, collisional ionization can effectively 
produce OVI at a temperature around $\log T\sim5.5$ while photoionization is effective 
only in low-density regions. The gas in clumps with large $N_{\rm HI}$ 
usually has low temperature and high density, making it difficult to 
produce large amounts of OVI ions. Neither HI nor OVI absorption 
is a good tracer of massive halos, such as the Coma cluster, because the ICM 
is too hot to produce these ions. Interestingly, one can 
see many filaments in both HI and OVI around the Coma cluster, 
which are produced by relatively cold gas being accreted by the cluster. 
This is consistent with results from the Ly$\alpha$ absorber survey 
of the Coma cluster, which found that Ly$\alpha$ absorption tends to
avoid the hot ICM \citep{Yoon2017}.

\begin{figure*}
    \centering
    \includegraphics[scale=0.6]{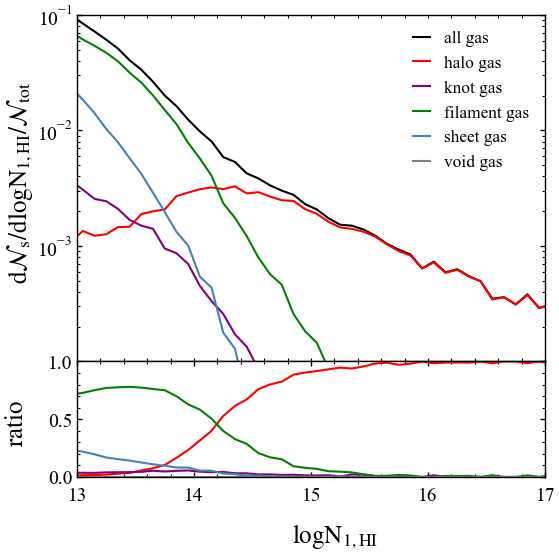}\includegraphics[scale=0.6]{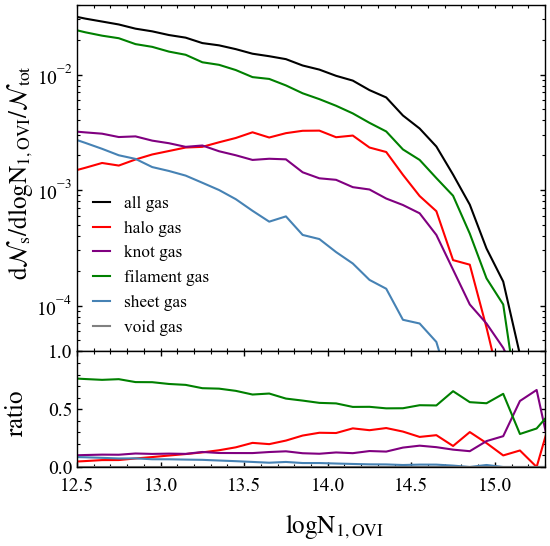}
	\caption{The upper panels show the distributions of HI (left) and OVI (right) column density obtained from the GW simulation. The column density is calculated by integrating the corresponding ion density 
	over an interval of $1\mpc$. Black lines show the results for all gas particles 
	regardless the types of the cosmic structure where they are located. 
	The color-coded lines show the contributions of different components of 
	the cosmic web. The contribution of the void gas is below the bottom of the plot. 
	The lower panels show the ratio of the distributions of different components of cosmic web to all gas.  
	See the text for details of the analysis. 
	}\label{fig_N1str}
\end{figure*}

More quantitative studies are required to understand which types of 
cosmic structure are traced well by these two types of absorption line systems.
Figure \ref{fig_N1str} shows the distribution of the column density in different 
cosmic structures in the GW simulation. The column density in this figure
is computed by integrating the density over an interval of $1\mpc$ along a line of sight, 
instead of over the whole HIR. This column density is referred to as $N_{1}$, 
and we sample the distribution of $N_1$ using a large number of random intervals. 
We classify each interval as halo, knot, filament, sheet or void, 
according to the structure type that dominates the column density of the 
interval. The distribution of the column density produced by a given type 
of the cosmic web is defined as $d{\cal N_{\rm s}}/d\log N_{1}/{\cal N_{\rm tot}}$, 
where $d{\cal N_{\rm s}}$ is the number of intervals with column density 
between $\log N_{1}$ and $\log N_{1}+d\log N_{1}$ for the type  
in question, while ${\cal N_{\rm tot}}$ is the total number of intervals 
regardless of structure types. We also show in the same figure the ratio of the distribution 
function between the gas in a given cosmic web type 
and the total gas.

As one can see, the total $N_{\rm 1, HI}$ distribution is roughly a power-law, 
similar to what is observed for quasar absorption line systems 
\cite[see e.g.][]{Danforth2016}. At $\log N_{\rm 1,HI}<14.2$, the $N_{\rm 1,HI}$ 
distribution becomes steeper. The distribution functions for different types of 
the cosmic web show that, at $\log N_{\rm 1,HI}>14.2$, the column density 
is dominated by gas in halos, while at $\log N_{\rm 1,HI}<14.2$ it 
is dominated by diffuse gas associated with filaments. The contribution from 
the other three components are negligible in this column density range. 
This is also seen in the column-density maps, where high column-density absorbers 
are compact and clumpy. \cite{Rudie2012ApJ, Rudie2013ApJ} found that HI absorption 
systems with $\log N_{\rm HI}>15$ are spatially associated with galaxies, 
broadly consistent with our results that they are dominated by gas associated with 
halos. 

The $N_{\rm 1, OVI}$ distribution is different from that for the HI. 
The total distribution is quite flat at $\log(N_{\rm 1, OVI})<14$ and declines 
rapidly at higher $N_{\rm 1, OVI}$, roughly consistent with observational 
results \citep[see e.g.][]{Danforth2016}. The decomposition into different 
cosmic web types clearly shows that the diffuse gas associated with filaments dominates 
over almost the entire column-density range plotted. More than 50\% of the OVI absorption 
is associated with filaments, and the fraction increases to about 80\% at 
the low $N_{\rm 1, OVI}$ end. Thus, OVI absorption is an promising probe 
of the diffuse gas component of the cosmic web. 
The COS-Halos project \cite[][]{Tumlinson2011ApJ,Tumlinson2013ApJ,Werk2012ApJS, Werk2013ApJS} detected 
many OVI absorption systems around galaxies in the local Universe. 
This is not in conflict with our results. The COS-Halos project 
purposely selected quasar sightlines that are close to local galaxies, 
while our results shown above are obtained from sightlines that are uniformly 
distributed in the sky. In fact, our results show that halo and knot gas 
also have important contributions to high column density systems with 
$\log(N_{\rm 1, OVI})>13.5$. This is consistent with the COS-Halos project, which 
found that most of the detected OVI absorption systems have $\log N_{\rm OVI}>13.5$.

We also do the same analysis for the other two simulations. 
The shapes of the column density distributions for both CM and LD are similar 
to those of the GW simulation. However, the amplitudes are smaller,  
as expected from the fact that the gas density in the GW region is higher than
those in the other two (see Table \ref{tab_HIR}). We also decompose 
the column density distribution into different cosmic web components. 
At $\log N_{1, \rm HI}>14.2$, the distribution is dominated by halo gas, while 
at $\log N_{1, \rm HI}<14.2$, filament gas is the most important 
contributor. For OVI, on the other hand, diffuse gas in 
filaments dominates the column-density distribution over the column-density  
range considered here. These results are qualitatively the same as those 
for the GW simulation, and are not shown here.
Note, however, that the cosmic web classification can be affected by the smoothing 
scale adopted in the calculation. To test the reliability of our results, we repeated our 
calculation using $R_{\rm s}=2\mpc$, instead of $0.5\mpc$, to make the 
classification. We also used different intervals to compute the column density. 
Our results are not affected significantly by these changes, and 
our basic conclusions remain the same.

We can also estimate the absorption column densities 
expected for real quasars in the three sky regions covered by our simulations.
Based on the SDSS DR14 quasar catalog, we find 682, 13711 and 6386 quasars with
$0.2 <z_{\rm AGN} <0.9$ in regions covered by GW, CM and LD, respectively.
As shown in \cite{Danforth2016}, quasars in this redshift range can be used 
to investigate their absorption line systems. For each real quasar, we integrate 
the ion densities along its sightline to obtain the HI and OVI column densities.
We find that there are 485, 8584, 4174 quasar sightlines with $\log N_{\rm HI}>13.5$ 
in the three regions respectively. Thus, more than 71, 63 and 65\% of the quasars in the
three regions are expected to have strong Ly$\alpha$ absorption lines. 
For a higher threshold of $\log N_{\rm HI} >14.2$, which selects absorption systems 
associated with halos, the numbers of quasar sightlines reduces to
128, 2150, and 784 in the three regions, respectively. For OVI,  
our simulations predicts 301, 4855, 1801 quasar sightlines
with $\log N_{\rm OVI}>13$ in their spectra, corresponding to 44, 35 and 28\%
of the quasars in the three regions, respectively.  
The fraction in the LD region is the lowest, while that in the GW the highest, 
as one expects given the difference in the mean mass density between the three 
regions. The numbers of quasar sightlines reduce to 
96, 1536, and 522, respectively, when a higher threshold of $\log N_{\rm OVI}>14$
is used. It should be pointed out, however, other factors can affect the detection 
of absorption lines. For example, quasar spectra with a high signal to noise ratio 
are needed for the detection of weak absorption lines. 
Detailed investigations are required to select proper targets to study 
quasar absorption line systems observationally.

As mentioned above, the column density distributions shown above 
cannot be compared directly with the observation. In the future, we will model quasar absorption line systems 
in more detail, using SPECEXBIN \citep{Oppenheimer2006MNRAS} to generate 
absorption profiles for various ions based on the predicted chemical abundance, 
ionization state, cosmic redshift and peculiar velocity, and to identify 
individual absorption lines \citep[see also][]{Dave2010,Huang2019MNRAS}. 
This will allow us to make direct comparisons with 
existing observational data in a statistical way. Furthermore, the results will also help us 
to identify quasars that are the most promising targets 
for studying the properties of the gas associated with different 
components of the cosmic web. Given the large number of quasars available,  
most of the massive filament structures in the three regions can be 
sampled quite densely by quasar sightlines. The constrained simulations 
can thus help to design observational strategies to maximize observational 
efficiency. In addition, the full information about the accretion, heating
and enrichment histories provided by the simulations will open a new avenue to 
interpret the observational data of individual quasar sightlines in terms 
of underlying physical processes that determine the properties of the 
absorption gas.

\section{Summary and discussion}\label{sec_sum}

Studying the distribution and properties of gas in the cosmic web is 
a key step towards understanding galaxy evolution and formation, and in searching for 
`missing baryons' that are not observed in galaxies. In this paper, we use 
constrained hydrodynamic simulations to study the properties and evolution of the
gas components in three representative regions of the local Universe, 
the Coma galaxy cluster, the SDSS great wall, and a large low density region at $z\sim0.05$. 
The initial conditions of the three simulations are reconstructed from the ELUCID 
project. We use these simulations to demonstrate how the gas associated with 
different components of the cosmic web can be studied using the
thermal Sunyaev-Zel’dovich (tSZ) effect, X-ray emission, and quasar absorption 
line spectra. Our main results can be summarized as follows.

Our simulations reveal that cluster, filament and low-density regions have very different
evolutionary histories. The Coma cluster experienced several violent merger 
events during the past six billion years, while filament and low-density regions evolve 
more slowly. Gas in the cosmic web is heated by both adiabatic compression and shocks 
induced by mergers and accretion. About half of the warm-hot intergalactic medium 
resides in filaments, and the fraction does not change significantly among the 
three simulated regions.

We find a strong and tight correlation between the local tidal field strength 
($\lambda_1$) and gas temperature ($T$). The relation consists of three 
power-law components: 
$\log T\propto  \log\lambda_1$ at $\log\lambda_1>1$;
$\log T\propto 2\log\lambda_1$ at $-0.5<\log\lambda_1<1$; and    
$\log T\propto {\rm const.}$ at $\log\lambda_1<-0.5$. 
This correlation, together with the tight correlation between the 
gas and total mass densities, can be used to predict the gas properties 
(density, temperature and pressure) associated with the cosmic web 
predicted by a pure dark matter simulation, thereby providing 
a way to empirically model the cosmic gas and to interpret observational data.

Our constrained simulations can reproduce the profiles of the Compton 
$y$-parameter and X-ray emission observed in the inner region of the Coma cluster. 
The simulations also produce, in maps of the $y$-parameter and X-ray,
some discontinuity features that are seen in real data, suggesting 
that our simulations capture well the heating processes associated with 
the merging history of the Coma cluster. Thus, the constrained simulations
provide a new avenue to understand how the gas in and around the Coma 
galaxy cluster was affected by its formation history and by the baryonic 
physics assumed in the simulation.  

We also investigate absorption line systems, such as Ly$\alpha$ and OVI, 
expected from our constrained simulations. 
Ly$\alpha$ absorption systems with $\log N_{\rm HI}>14.2$ are mostly 
associated with gas within halos, while most of the OVI absorption and low-$N_{\rm HI}$ absorption 
are produced by gas in filaments. OVI absorption lines are mainly connected to 
massive filaments, and small filaments, in particular those in low-density regions, 
produce Ly$\alpha$ absorption systems with low column densities. 
Combined with quasars observed behind the three simulated regions, our constrained 
simulations can be used to select promising quasar candidates to 
explore the absorption of the gas associated with specific components of the 
cosmic web, e.g. massive filaments,  thereby helping to understand how  
the spatial distribution and properties of the gas in the cosmic web 
is affected by formation processes and feedback processes.

The simulations presented here do not include AGN feedback. In the future, 
we will perform hydrodynamic simulations with different implementations of AGN 
feedback. As shown in previous studies using random phase simulations,   
AGN feedback can affect the gas component in the cosmic web in its thermodynamic and 
chemical properties, as well as its spatial distribution. Such feedback may thus  
significantly affect the predicted SZ, X-ray, and absorption-line properties of 
the cosmic web. Detailed comparisons between constrained simulations implementing different 
models of feedback and observational data in the same cosmic web thus provide 
an important avenue to discriminate feedback models.

\section*{Acknowledgements}
We thank M. S. Mirakhor for kindly providing the observational data to us. We thank the referee for the useful report.
This work is supported by the National Key R\&D Program of China (grant No. 2018YFA0404503), the National Natural Science Foundation of China (NSFC, Nos.  11733004, 12192224, 11890693, 11421303, 11833005, 11890692, 11621303),  and the Fundamental Research Funds for the Central Universities. We acknowledge the science research grants from the China Manned Space Project with No.
CMS-CSST-2021-A03. The authors gratefully acknowledge the support of Cyrus Chun Ying Tang Foundations. We further acknowledges the science research grants from the China Manned Space Project with NO. CMS-CSST-2021-A01 and CMS-CSST-2021-B01. 
WC is supported by the STFC AGP Grant ST/V000594/1 and Atracci\'{o}n de Talento Contract no. 2020-T1/TIC-19882 granted by the Comunidad de Madrid in Spain.
The work is also supported by the Supercomputer Center of University of Science and Technology of China.

\bibliography{ref.bib}
\appendix

In this paper, we adopted two methods to obtain SZ $y$-maps and X-ray brightness 
maps (Sections \ref{sec_tsz} and \ref{sec_xray}).
The SphM method, in which gas properties are calculated on small cells 
with sizes less than the gravitational softening length using the SPH smoothing kernel, 
is the standard technique to obtain fluid quantities from SPH particles.
However, this technique is very time consuming for a large simulation volume.  
The other method, ParM, which directly uses the properties of gas particles to estimate 
quantities on a given pixel, is much faster. However, the ParM method is 
valid only when the pixel size is larger than the SPH smoothing length. 
It is thus less accurate in lower density regions. 

\begin{figure*}
    \centering
    \includegraphics[scale=0.4]{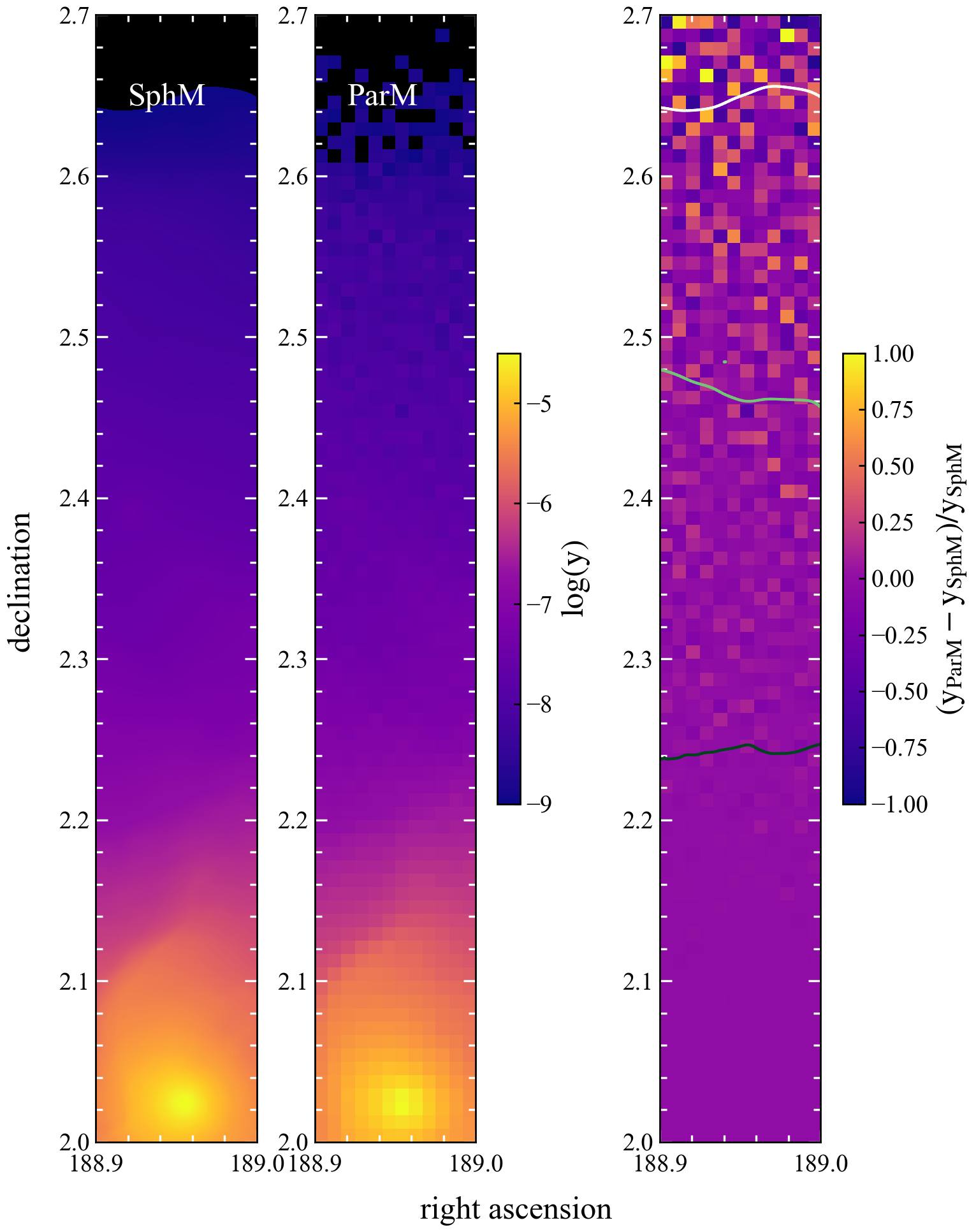} \includegraphics[scale=0.4]{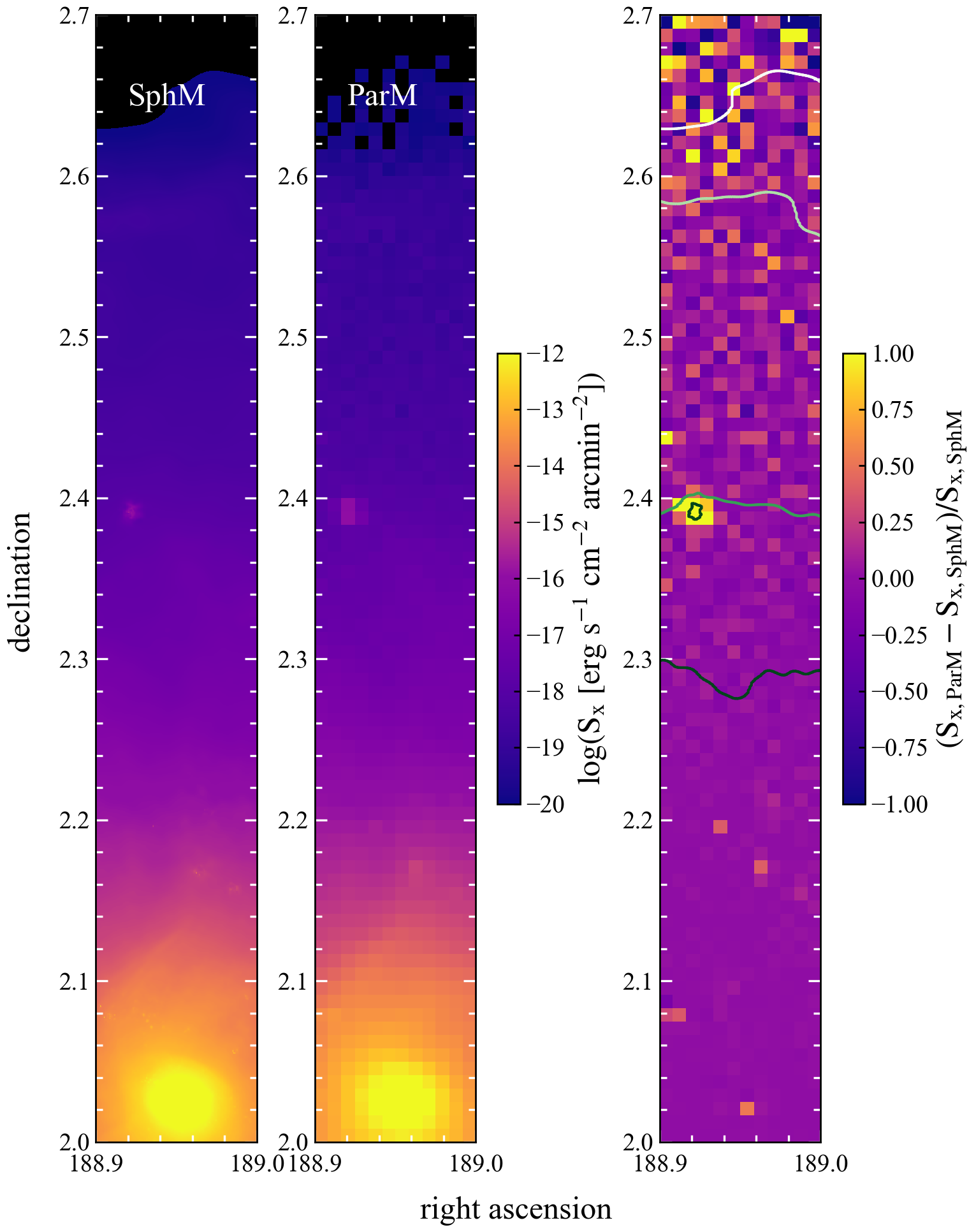}
	\caption{The left three panels show SZ $y$-maps of a small region in the GW simulation. 
	The left one is obtained by using the SphM method, the middle one by using ParM, 
	and the right one shows the difference between the two methods. The three contours in 
	the difference map show $\log y_{\rm SphM}= -7$, -8 and -9, respectively.
	The right three panels show the results for the X-ray brightness, 
	$S_{\rm x, SphM}$, $S_{\rm x, ParM}$ and their difference, respectively.
	The contours in the right panel show $\log S_{\rm x,SphM}= -17$, -18, -19 and -20, respectively.}\label{fig_compare}
\end{figure*}

In this Appendix, we compare the results obtained using the two methods.  
Here we select a small region in the GW simulation for the comparison. 
This region covers a large dynamical range in both $y$ and $S_{\rm x}$, 
which is ideal for testing the two methods. The maps of the two quantities 
obtained using the two methods, together with their fractional differences, 
are shown in Figure \ref{fig_compare}. 
Note that the maps for the ParM method use pixels of 0.5 arcmin, the
same as that shown in Figure \ref{Fig_gwxrayYmap}, while the SphM maps 
have a much higher resolution. In calculating the difference maps, 
we use a pixel size of 0.5 arcmin for both SphM and ParM results. 

\begin{figure*}
    \centering
    \includegraphics[scale=0.6]{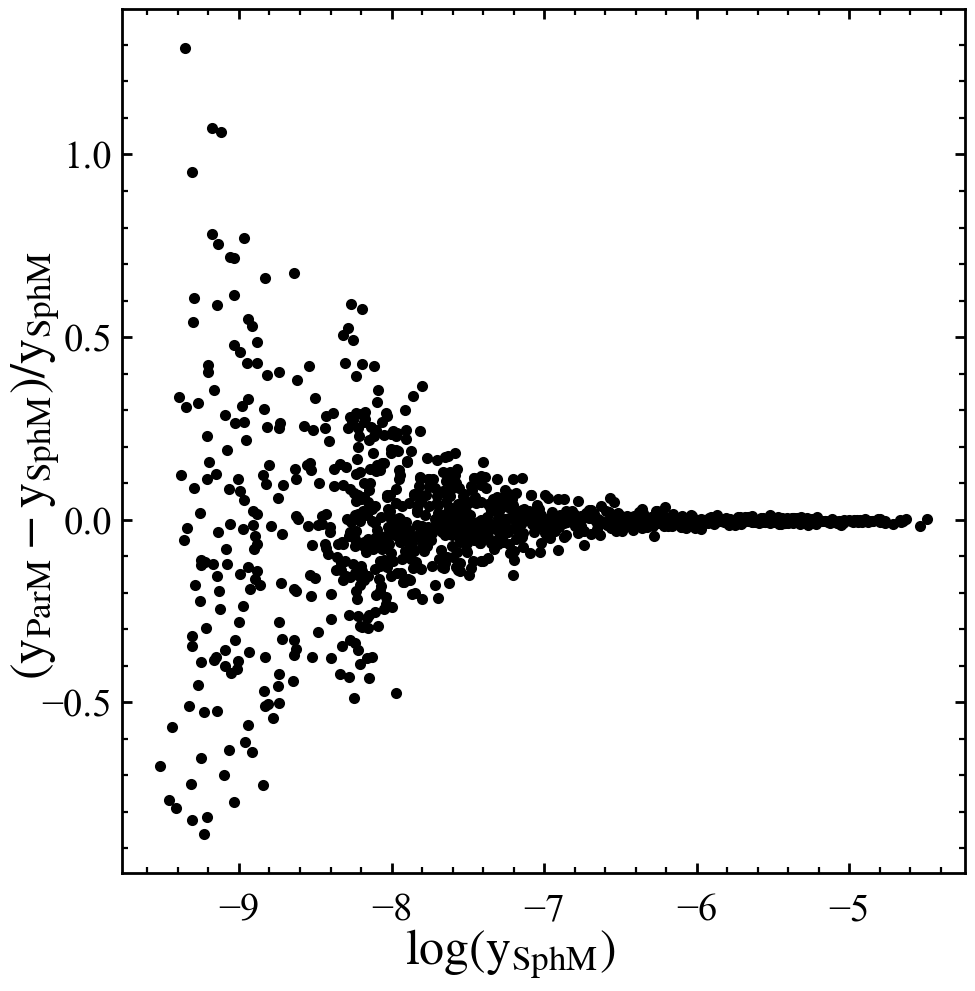} \includegraphics[scale=0.6]{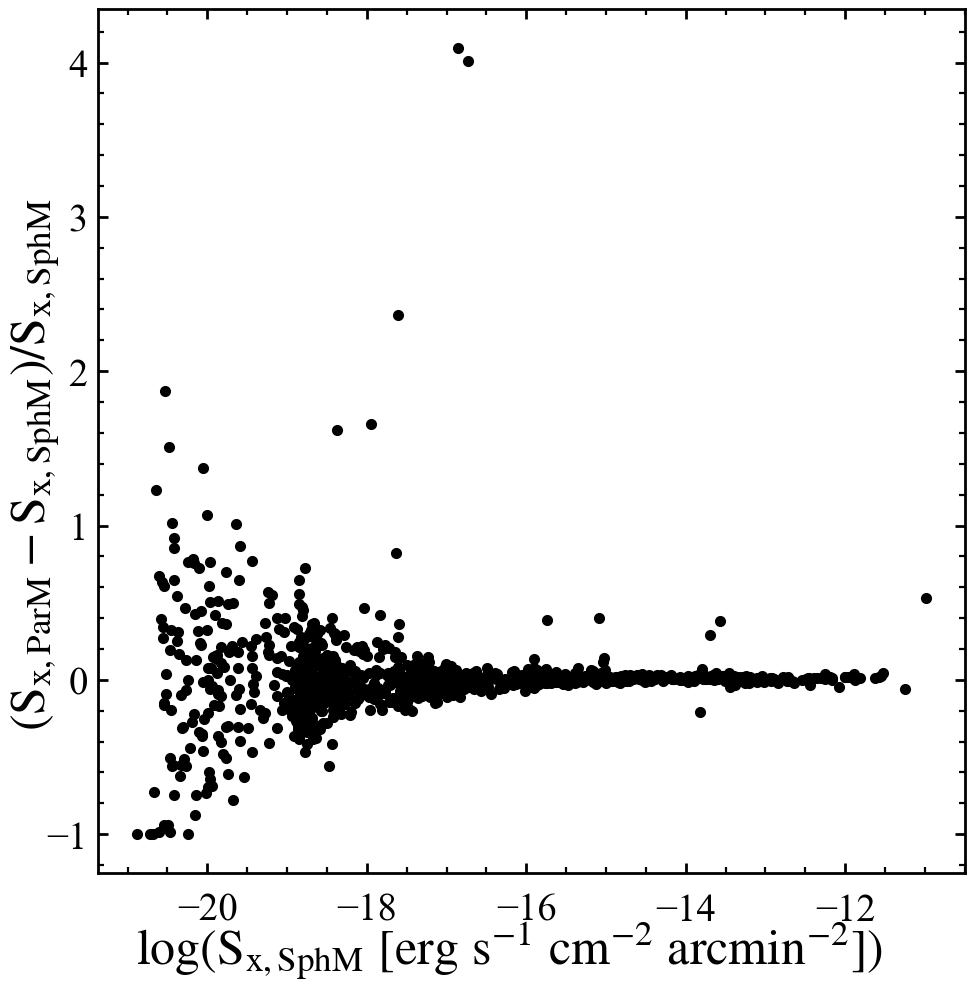}
	\caption{The left panel shows $(y_{\rm ParM}-y_{\rm SphM})/y_{\rm SphM}$ as a function of $\log y_{\rm SphM}$ for all pixels in the sky region shown in Figure \ref{fig_compare}. The right panel shows $(S_{\rm x,ParM}-S_{\rm x,SphM})/S_{\rm x,SphM}$ as a function of $\log S_{\rm x,SphM}$ for all pixels in the same sky region.
	}\label{fig_sta}
\end{figure*}

As one can see, for both $y$ and $S_{\rm x}$, the two methods 
give almost identical results when their values are high.  
The differences become larger as $y$ and $S_{\rm x}$ decrease. 
This is expected as the ParM method is not valid for low density regions 
where the SPH smoothing kernel is larger than the pixel size 
and where both $y$ and $S_{\rm x}$ are expected to be low. 
In Figure \ref{fig_sta}, we also show the difference between the two methods
as a function of $y_{\rm SphM}$ and $S_{\rm x, SphM}$, respectively. 
The fractional differences in $y$ and $S_X$
are less than 0.5 at $\log y_{\rm SphM}> -8.5$
and $\log S_{\rm x, SphM}> -19$, respectively, so that the results obtained 
from ParM are reliable for regions above these thresholds.
A small number of pixels have $\log S_{\rm x, SphM}> -19$ 
and yet a large difference between the two methods. 
As shown in Figure \ref{fig_compare}, these pixels are associated  
with the poor performance of ParM in handling small compact structures with 
sizes comparable to the pixel size. 
In general, the difference between ParM and SphM becomes smaller 
when the comparison uses a larger pixel size.

\nolinenumbers
\clearpage
\end{document}